\pdfoutput=1
\begingroup\expandafter\expandafter\expandafter\endgroup
\expandafter\ifx\csname pdfsuppresswarningpagegroup\endcsname\relax
\else
\fi

\documentclass[preprint]{elsarticle} 

\usepackage[utf8]{inputenc}
\usepackage[T1]{fontenc}
\usepackage[english]{babel}
\usepackage{textcomp}
\usepackage{booktabs,multicol}

\usepackage[hyphens]{url}
\biboptions{sort&compress, square, comma}
\bibpunct{{\unskip~[}}{]}{,}{n}{}{;} 

\usepackage[breaklinks=true, linkcolor=blue, citecolor=blue, colorlinks=true]{hyperref}

\usepackage{latexsym, amsmath, amssymb}
\usepackage{graphicx, wrapfig, subcaption}
\usepackage{color}
\usepackage[pdftex,usenames,dvipsnames,changebar]{xcolor}
\usepackage[xcolor]{changebar}
\usepackage{listings}
\usepackage{nicefrac}
\usepackage[shortlabels]{enumitem}
\usepackage[binary-units]{siunitx}

\journal{Journal of Supercomputing}

\definecolor{dkgreen}{rgb}{0,0.6,0}
\definecolor{gray}{rgb}{0.5,0.5,0.5}
\definecolor{mauve}{rgb}{0.58,0,0.82}

\def\CC{{C\nolinebreak[4]\hspace{-.05em}\raisebox{.4ex}{\footnotesize ++}}}

\usepackage[normalem]{ulem}

\def\WCPU{Intel Xeon E5-2630 v3}
\def\WGPU{Nvidia Tesla K40c}
\def\CCPU{Intel Xeon E5-2660 v3}
\def\CGPU{Nvidia Tesla K40m}
\def\UpT{Up Triangle}
\def\DownT{Down Triangle}
\def\Diam{Diamond}
\def\Comm{Communication}
\newcommand{\tx}[1]{$#1\times$}

\begin{document}

\begin{frontmatter}

\title{Applying the swept rule for solving explicit partial differential equations on heterogeneous computing systems}

\author[osu]{Daniel J.~Magee\fnref{fn1}}
\author[osu]{Anthony S.~Walker}
\author[osu]{Kyle E.~Niemeyer\corref{cor1}}
\ead{kyle.niemeyer@oregonstate.edu}

\address[osu]{School of Mechanical, Industrial, and Manufacturing Engineering, Oregon State University, Corvallis, OR 97331, USA}

\fntext[fn1]{Current address: Los Alamos National Laboratory, Los Alamos, New Mexico 87545, USA}
\cortext[cor1]{Corresponding author}

\begin{abstract}
Applications that exploit the architectural details of high-performance 
computing (HPC) systems have become increasingly invaluable in academia 
and industry over the past two decades.
The most important hardware development of the last decade in HPC has 
been the General Purpose Graphics Processing Unit (GPGPU), 
a class of massively parallel devices that
now contributes the majority of computational power in the top 500 supercomputers.
As these systems grow, small costs such as latency---due to the fixed cost of 
memory accesses and communication---accumulate
in a large simulation and become a significant barrier to performance.
The swept time-space decomposition rule is a communication-avoiding technique for
time-stepping stencil update formulas that attempts to reduce latency costs.
This work extends the swept rule by targeting heterogeneous, 
CPU\slash GPU architectures representing current and future HPC systems. 
We compare our approach to a naive decomposition scheme with two test 
equations using an MPI+CUDA pattern on 40 processes
over two nodes containing one GPU. 
The swept rule produces a factor of 1.9 to 23
speedup for the heat equation and a factor of 
1.1 to 2.0 speedup for the Euler equations, 
using the same processors and work distribution, and with the best possible configurations.
These results show the potential effectiveness of the swept rule for different equations and numerical schemes on massively parallel compute systems that incur substantial latency costs.
\end{abstract}

\begin{keyword}
Domain decomposition \sep Heterogeneous computing \sep partial differential equations \sep computational fluid dynamics \sep communication-avoiding algorithms
\end{keyword}


\end{frontmatter}

\section{Introduction}

Computational fluid dynamics (CFD) simulations lie at the heart of technological development in industries vital to high and rising standards of living around the world.
However, performing simulations at the level of fidelity necessary for continuous insight
consumes more resources than individual workstations can reasonably accommodate.
As a result, these simulations are typically performed on high-performance computing (HPC)
systems, distributing problems across many nodes that each contain multiple multicore
central processing units (CPUs), and increasingly in combination with other specialized ``accelerator'' co-processors. 
These heterogeneous---that is, containing more than one processing architecture---computing systems have become ubiquitous in areas of research that depend on large amounts of data, complex numerical transformations, or densely connected systems of constraints.
Steady progress in addressing these problems requires developing algorithms that consider hardware capabilities (e.g., computational intensity) and limitations (e.g., bandwidth\slash communication).

In many ways recent improvements in computational capacity have been sustained by the development of accelerators or co-processors, such as general purpose graphics processing units (GPGPUs) or the Intel Xeon Phi manycore processor, that augment the computational capabilities of the CPU. These devices have grown in power and complexity over the last two decades, leading to an increasing reliance on them for enabling efficient floating-point computation on HPC systems~\cite{ALEXANDROV20161}.
Latency and bandwidth costs limit the performance of applications, such as CFD simulations, that require inter-node communications as the system grows in complexity, computational power, and physical size.
Bandwidth is the amount of memory that can be communicated per unit of time, and latency is the fixed cost of a communication event: the travel time of the leading bit in a message. 

Solving partial differential equations (PDEs) on HPC systems using explicit numerical methods requires domain decomposition (a heuristic for dividing the computational domain across the processors), requiring inter-node communication of small data packets for boundary information at every time step. 
The frequency of these communication events renders their fixed cost, latency, a significant barrier to the performance of these simulations.

Our work is aligned with the overall goals of the HPC development community and seeks to address, however nascently, two of the challenges on the route to exascale computing systems recently identified by Alexandrov: the need for ``novel mathematical methods\dots{} to hide network and memory latency, have very high computation\slash communication overlap, have minimal communication, have fewer synchronization points'', and ``mathematical methods developed and corresponding scientific algorithms need to match these architectures [standard processors and GPGPUs] to extract the most performance. This includes different system-specific levels of parallelism as well as co-scheduling of computation''~\cite{ALEXANDROV20161}.

In this article we describe the development and performance analysis of a PDE solver targeting
heterogeneous computing systems (i.e., CPU\slash GPU) using the swept rule: 
a communication-avoiding, latency-hiding domain decomposition scheme~\cite{alhubail:16jcp,Alhubail:2016arxiv}. 
Section~\ref{sec:hRwork} describes recent work on domain decomposition schemes with particular attention to applications involving PDEs and heterogeneous systems. Section~\ref{sec:obj1} lists the questions this study seeks to answer. 
Section~\ref{sec:hMethods} introduces swept time-space decomposition and discusses the experimental hardware, procedure, and factors used to evaluate performance. 
In Section~\ref{sec:hResults} we present the results of the tests and describe the hardware and the testing procedures used; lastly in Section~\ref{sec:hConc} we draw further conclusions, describe future challenges, and outline plans for prioritizing and overcoming them.

\section{Related work}
\label{sec:hRwork}

The swept rule is described here from a high level and supported by a more detailed description in Section~\ref{sec:hSweptDecomp}. 
It is a latency reduction technique originally developed by Alhubail et al.~\cite{alhubail:16jcp} to solve PDEs on large-scale computing systems where the cost of communication can be substantial. 
The swept rule works by prioritizing solution of all possible spatial locations with local information; points that cannot be solved are skipped for the time being. 
These ``skipped'' points lie on boundaries where advancing the solution in time would require inter-node communication.
The newly calculated local information is then used to do the same for the next step.
However, even more boundary points are now skipped because domain boundary information was not communicated. 
Steps are taken in this manner until communication is necessary to proceed further.
When communication is necessary, the needed boundary points to solve the skipped points are communicated in one step and computation resumes. 
By compressing all communication into a single step, this algorithm reduces the cost of communication.

This method---the swept time-space decomposition rule---was first developed for one-dimensional problems on a CPU-based system by Alhubail et al.~\cite{MaithamRepo,alhubail:16jcp}. 
In addition, Alhubail and Wang applied this procedure to automatically generate C source code for solving the heat and Kuramoto--Sivashinsky equations using the swept rule on CPU-based systems~\cite{AIAAWang}. 
They later extended this work to two dimensions \cite{Alhubail:2016arxiv}.
Wang also showed how complex numerical schemes can be decomposed into ``atomic'' update formulas, a series of steps requiring only a three-point stencil, suitable for the swept rule~\cite{WangDecomp}. 

In our previous work~\cite{OurJCP}, we investigated methods for exploiting the memory hierarchy on a single GPU in a swept time-space solver for stencil computations. 
We use this technique, which we refer to as ``\texttt{lengthening}'', in the implementation of the swept rule discussed here and contrast it with another method for dealing with complex schemes, ``\texttt{flattening}'', which we used in our previous GPU-only study~\cite{OurJCP}. Section~\ref{sec:hPrimaryData} quantitatively compares the two techniques. 
These articles comprise the body of work on the swept rule to date, upon which this paper expands. 
Related works consist of mostly parallel-in-time methods and communication-avoiding algorithms, but other studies on numerical stencils and memory use are also relevant.

Memory hierarchies are defined by a series of locations where memory is scarce and fast, to where it is plentiful and slow. 
By working on data in the limited fast-memory space as long as possible, communication-avoiding algorithms accelerate computations by reducing
inter-process communication or accesses to global memory in parallel programs.
Swept time-space decomposition is a type of communication-avoiding algorithm because it seeks to reduce the number of communication events between the processor and less-accessible memory resources. 
Unlike most communication-avoiding algorithms, it does not perform redundant operations. 
The heterogeneous communication-avoiding LU factorization algorithm presented by Baboulin et al.~\cite{BABOULIN201217} splits the tasks between the GPU and CPU and minimizes inter-device communication. 
Their results show an appreciable benefit from splitting the types of tasks performed on the CPU and GPU, which reduces overall communication and effectively overlaps computation and communication.
Demmel et al.~\cite{demmel2008avoiding} developed communication-avoiding Krylov subspace methods, and
Ballard et al.~\cite{ballard2011minimizing} minimized communications in linear algebra techniques such as LU and QR factorization.
Khabou et al.~\cite{khabou2013lu} developed a communication-avoiding LU\_PRRP factorization method, and Solomonik et al.~\cite{solomonik2017communication} produced a communication-avoiding algorithm for solving symmetric eigenvalue problems.
The primary difference between the swept rule and these examples is that they are focused on specific linear algebra applications rather than solving PDEs.

Swept time-space decomposition is also conceptually related to parallel-in-time methods~\cite{Gander2015}, such as multigrid-reduction-in-time~\cite{falgout2014parallel}. 
These algorithms overcome the interdependence of solutions in the time domain to parallelize the problem as if spatial. 
This class of techniques iterates over a series of fine and coarse grids using an initial guess for the entire solution domain and effectively smooths out the errors in the solution. 
Historically, parallel-in-time methods were considered unsuitable for nonlinear problems since the use of coarse grids degraded efficiency and accuracy~\cite{alhubail:16jcp}. 

However, recent developments applying optimization and auto-tuning techniques have matched the scaling of linear solvers~\cite{MGRITImprove}. 
Parareal---developed by Lions et al.~\cite{lions2001resolution}---is a parallel-in-time method that solves multiple time steps in parallel on a fine grid and corrects the results on a coarse grid until the solution converges, resulting in a solution with the accuracy of the fine grid. 
Wu and Zhou proposed a new local time-integrator for this method that shows considerable promise for accelerating convergence rates in fractional differential equations~\cite{PararealWu}.

Other such examples of parallel-in-time methods include PFASST developed by Emmett and Minion~\cite{emmett2012toward}, and interwoven PFASST and Parallel Multigrid from Minion et al.~\cite{minion2015interweaving}. 
Gander and Guttel~\cite{gander2013paraexp} developed PARAEXP---a parallel integrator for linear IVPs---and Gander and Neumuller~\cite{gander2016analysis} later presented a new technique for parabolic problems. 
These are some of the methods developed for parallel-in-time integration but by no means a comprehensive list of studies. 

Studies of stencil optimization techniques over the last decade often address concerns closely related to the work presented here.
Datta et al.~\cite{VolkovDatta2008} explored domain decomposition with 
various launch parameters on heterogeneous architectures and
nested domain decomposition within levels of the memory hierarchy. 
Malas et al.~\cite{MalasHager} previously explored similar diamond tiling 
methods by using the data dependency of the grid to improve cache use. 

Though the community has made progress in algorithms designed to reduce 
communication costs in distributed systems, such distributed, remote, multi-node systems have become increasingly heterogeneous in recent years.
As a result, implementing CFD codes effectively on these systems has become more complex. 
Furthermore, as distributed-memory HPC systems continue to grow in size, latency between nodes will continue to impede time-to-solution. 
As a result, domain decomposition on these systems has received a good deal of recent attention. 
For example, Huerta et al.~\cite{DOEbenchmarks} used methods from process engineering, including experimental design and non-continuous linear models in an experimental parameter space paradigm, to investigate the performance of a well-known workload division benchmark used to rank HPC clusters on a heterogeneous system. 
This technique shows considerable promise for future studies of the swept rule with a more-mature code base. 
However, at our current stage, such a detailed analysis would not provide actionable insights beyond what we have already gleaned from our comparatively simpler methods.

\section{Objectives} \label{sec:obj1}

This study applies swept time-space decomposition to explicit stencil computations intended for distributed-memory systems with heterogeneous architecture, that is, systems with one or more CPUs combined with GPU co-processors. 
In this study we use two systems with multiple CPU cores and an Nvidia GPU.
The software implementing the swept scheme for heterogeneous systems, \texttt{hSweep}, is written in \CC{}\slash CUDA; it uses the Message Passing Interface (MPI) library~\cite{Clarke1994} to communicate between CPU processes and the CUDA API to communicate between GPU and CPU. 
As described in \ref{app:material} this software is openly available, and we archived the version used to produce the results in this study.

While stencil computation is a relatively simple procedure, i.e., applying linear operations to points on a grid, the complexities introduced by computing-system heterogeneity and swept time-space decomposition require a significant number of design decisions. 
In this work we investigated the performance impact of the most immediately salient and configurable decisions and constrained other potential variations with reasonable or previously investigated values. 
The primary focus of this paper to determine if the swept rule reduces the simulation run time using the most favorable launch configurations over a large range of grid sizes.
This raises several sub-questions that our study considered: \begin{enumerate}[nosep,topsep=0pt]
 \item How should we organize the stencil update formula for multi-step methods
 (discussed in Section~\ref{sec:hPrimaryData})?
 \item How much work should we give to the GPU in a heterogeneous system?
 \item Does the size of the domain of dependence substantially affect performance?
\end{enumerate}

\section{Methodology}
\label{sec:hMethods}

\subsection{Swept decomposition}
\label{sec:hSweptDecomp}

The swept rule exhausts the domain of dependence---the portion of the space-time grid that can be solved given a set of initial values, referred to here as a ``block''---before passing the grid points on the borders of each process. 
We refer to the program that implements the swept rule as \texttt{Swept} and to the program that uses naive domain decomposition (i.e., that passes all boundary between processes at each time step) as \texttt{Classic}\footnote{Alhubail and Wang~\cite{alhubail:16jcp} use the term ``straight'' decomposition where we use \texttt{Classic}.}.
Using the swept rule, the simulation may continue until no spatial points have available stencils; the required values may then be passed to the neighboring process (i.e., neighboring subdomain) in a single communication event. 

\begin{figure}[htbp]
  \centering
  \begin{subfigure}[t]{0.46\textwidth}
    \includegraphics[scale=1,width=\textwidth,trim={1.8cm 2.2cm 2.2cm 0.75cm},clip]{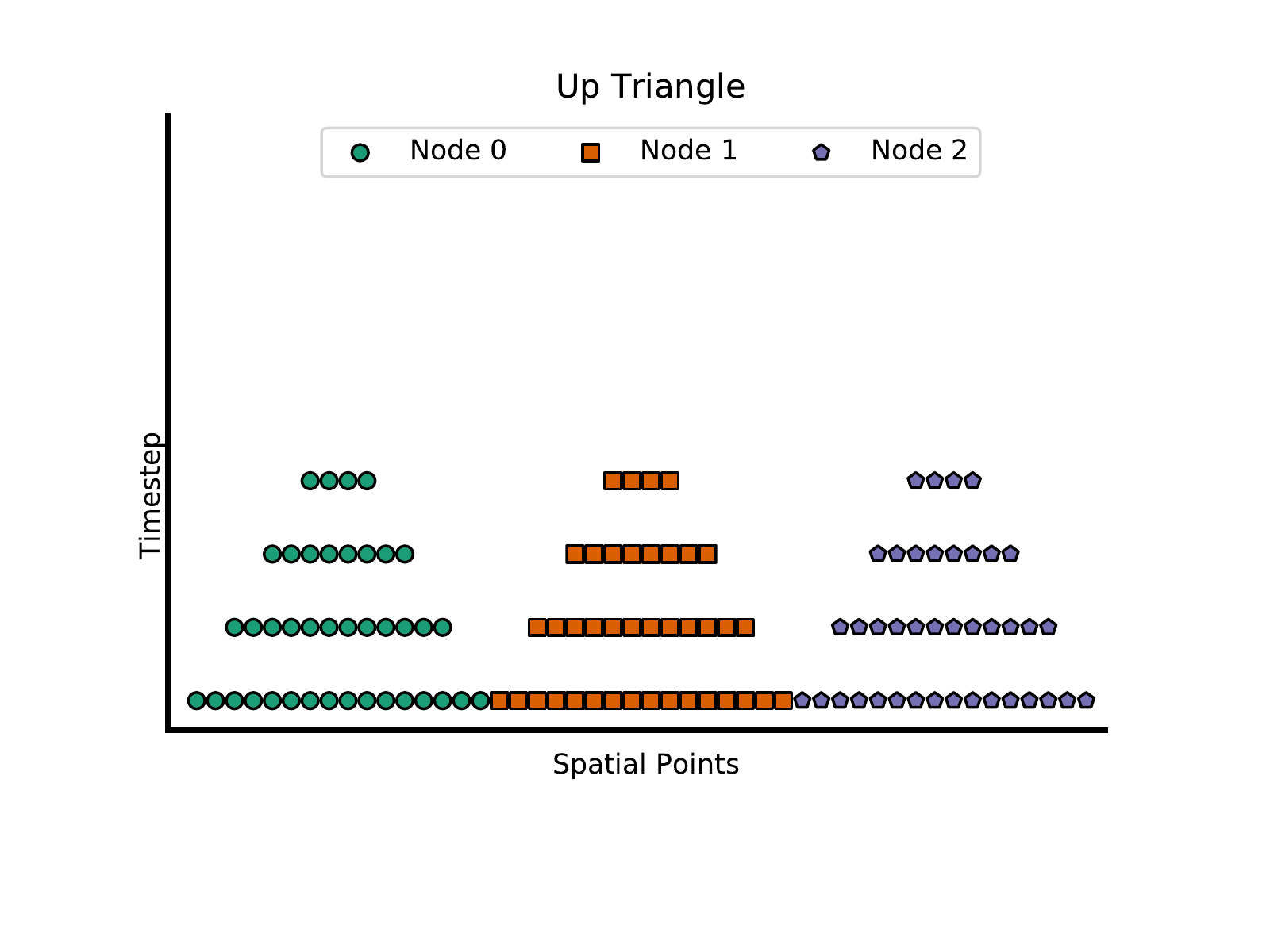}
    \caption{The first phase of the swept rule.}
    \label{fig:upTriangle}
  \end{subfigure}
  ~~
  \begin{subfigure}[t]{0.46\textwidth}
    \includegraphics[scale=1,width=\textwidth,trim={1.8cm 2.2cm 2.2cm 0.75cm},clip]{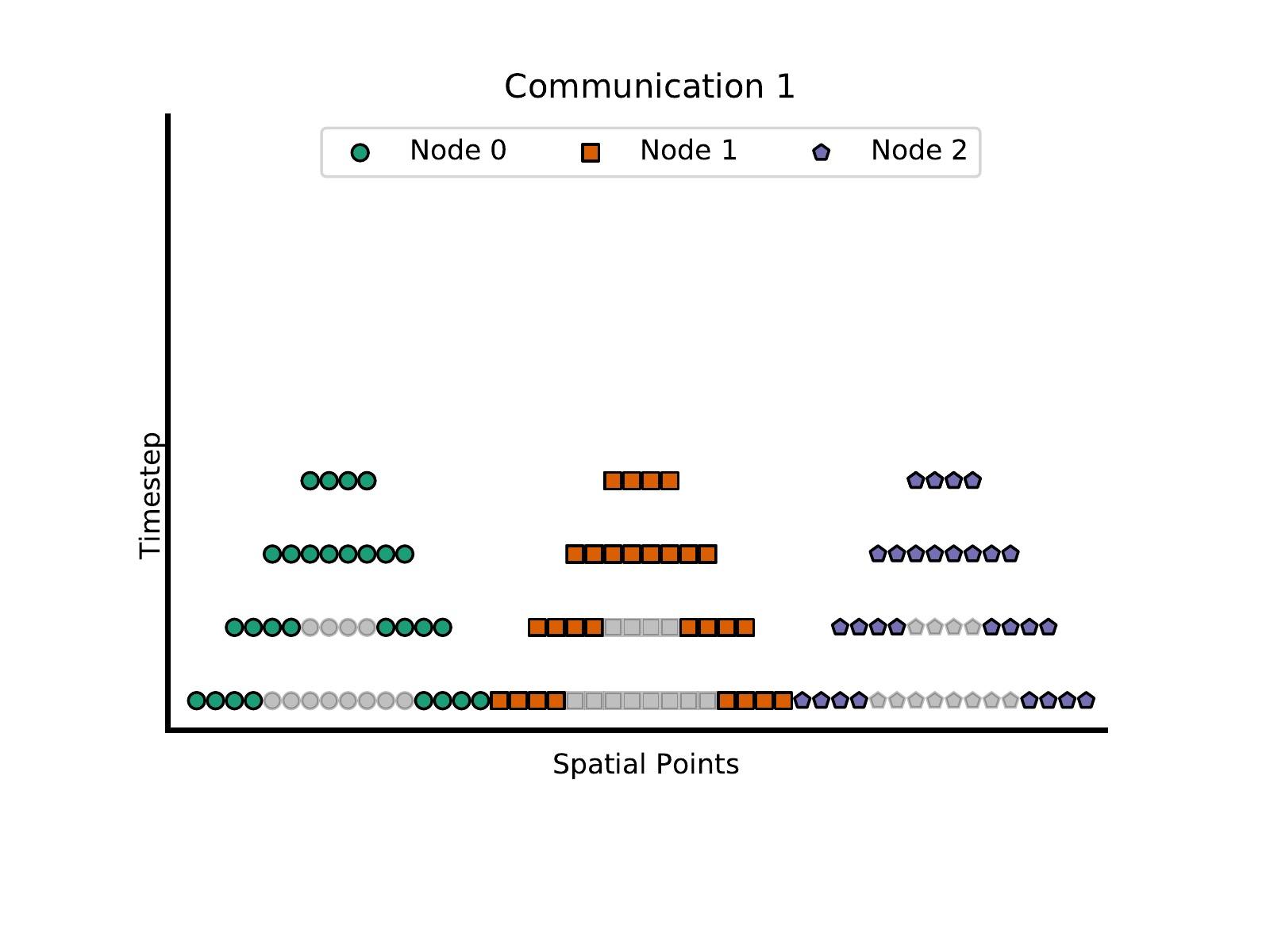}
    \caption{The first communication phase.}
    \label{fig:comm1}
  \end{subfigure}
  ~~
  \begin{subfigure}[t]{0.46\textwidth}
    \includegraphics[scale=1,width=\textwidth,trim={1.8cm 2.2cm 2.2cm 0.75cm},clip]{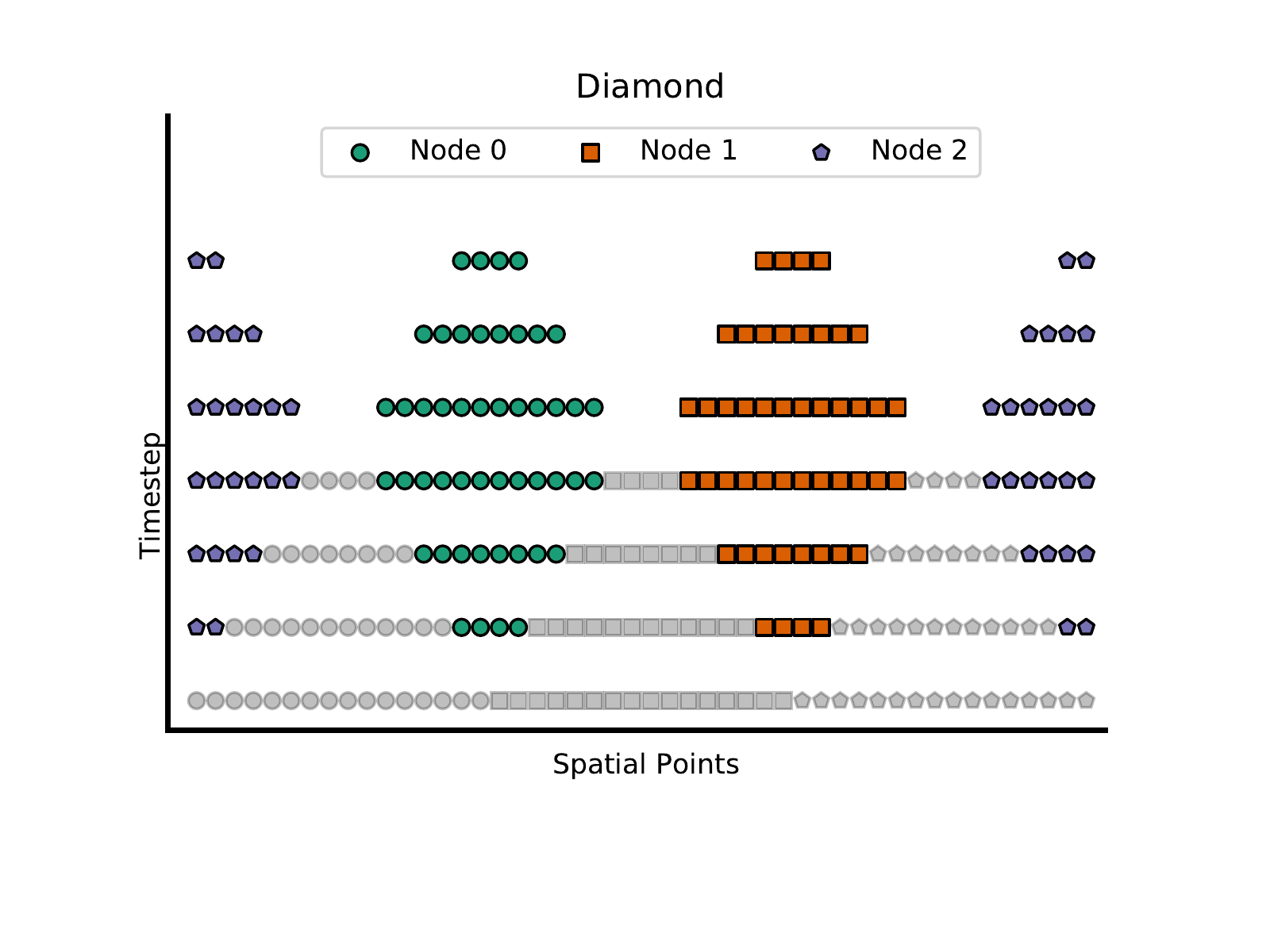}
    \caption{The diamond phase of the swept rule.}
    \label{fig:diamond}
  \end{subfigure}
  ~~
  \begin{subfigure}[t]{0.46\textwidth}
    \includegraphics[scale=1,width=\textwidth,trim={1.8cm 2.2cm 2.2cm 0.75cm},clip]{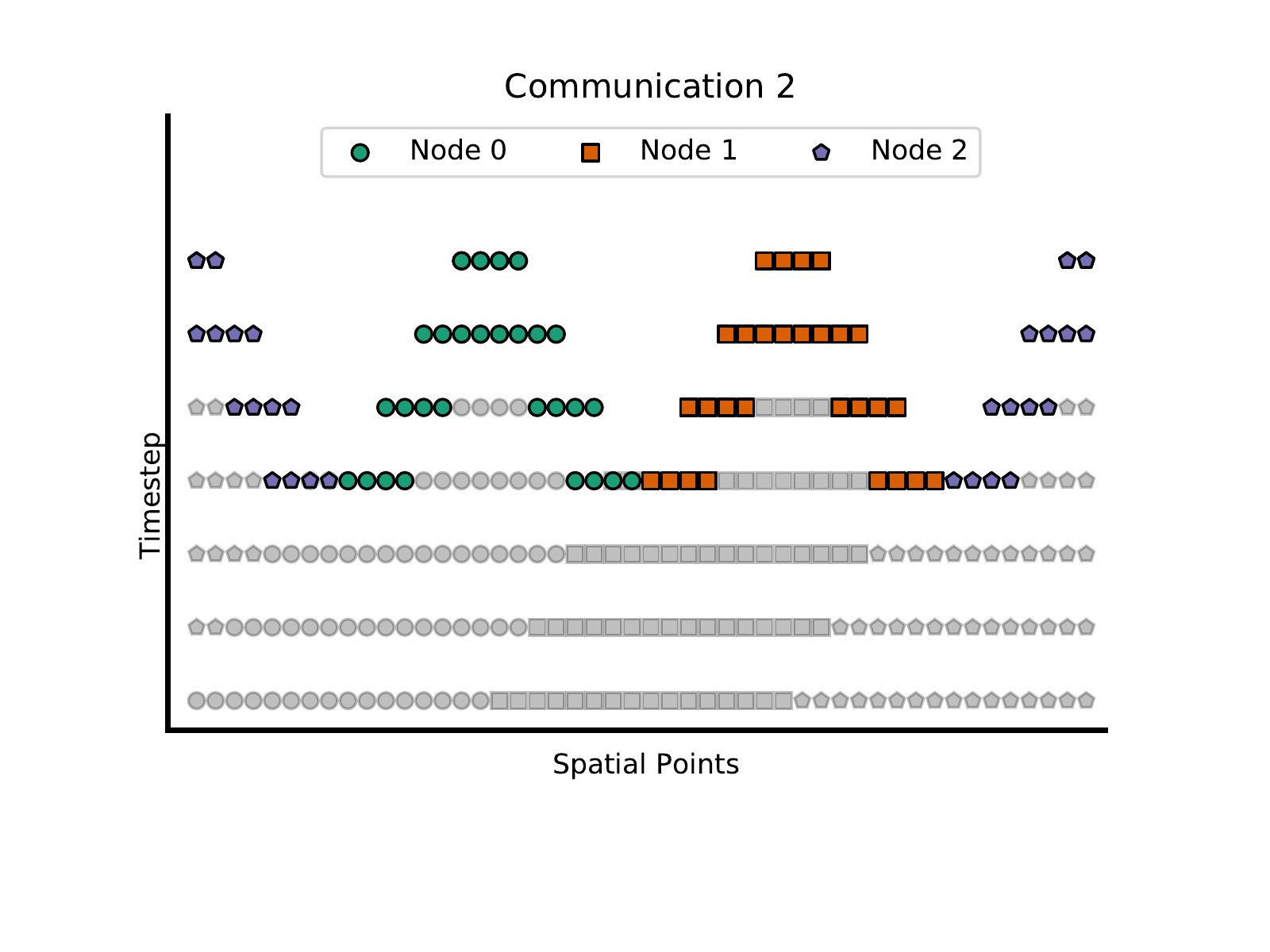}
    \caption{The second communication phase.}
    \label{fig:comm2}
  \end{subfigure}
  ~~
  \begin{subfigure}[t]{0.46\textwidth}
    \includegraphics[scale=1,width=\textwidth,trim={1.8cm 2.2cm 2.2cm 0.75cm},clip]{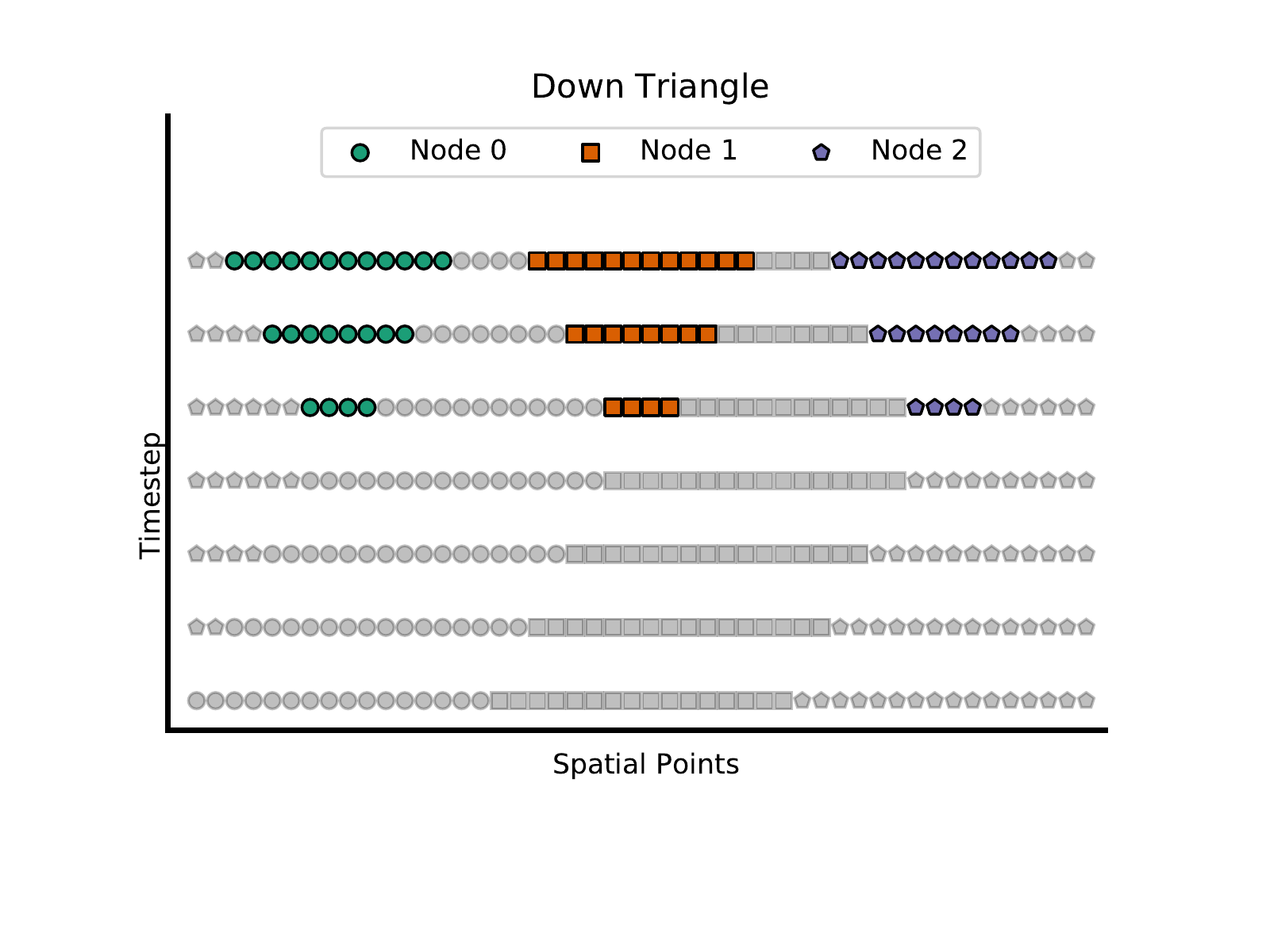}
    \caption{The down triangle phase of the swept rule.}
    \label{fig:downTriangle}
  \end{subfigure}
  
 \caption{The stages of the swept process with a five-point stencil.}
 \label{fig:sweptProcess}
\end{figure}

The swept rule in one spatial dimension can be broken down into four major steps: \UpT{}, \Diam{}, \DownT{}, and \Comm{}. 
Figure~\ref{fig:sweptProcess} shows this process for a five-point numerical stencil using three computing nodes.
The swept progression depends on the numerical discretization used regardless of the problem.
For example, here we solve Euler's equations with a five-point central difference, but solving the Navier--Stokes equations would progress in the same manner if solved with a five-point central difference.

Figure~\ref{fig:upTriangle} shows the first phase of the process (\UpT{}), which is created by removing twice the necessary number of boundary points from each subsequent time step which results in a triangular shape. 
After stepping in time to when no additional points be be advanced, the first \Comm{} step occurs. 
Shown in Figure~\ref{fig:comm1}, the highlighted points are the only points necessary to continue the computation. 
So, these points are passed to the neighboring nodes in one operation to allow the computation to continue. 

As you can see in Figures~\ref{fig:upTriangle} and \ref{fig:comm1}, the \UpT{} and \Comm{} steps leave a void between the subdomains of each computing node. 
This leads to the next step (\Diam{}) that fills the remaining void and builds another \UpT{} on top of it. 
This progresses similarly to the first \UpT{} phase by adding or removing twice the number of boundary points in each time step as is needed to fill the voided and build the next part. 

The next communication event then occurs as shown in Figure~\ref{fig:comm2}, and computation cycles back into the original position for the next phase. 
The \Diam{} and \Comm{} phases can be repeated as many times as is necessary to reach the desired time step. The computation is completed by the \DownT{} step shown in Figure~\ref{fig:downTriangle}.

This simple example provides a visual understanding of the process, but the swept rule is not limited to this specific case. 
This process can performed on a node and\slash or system level, e.g., each node may have a GPU and CPU which can add a second level of communication reduction depending on the implementation strategy. 
It can also depend on the stencil update formula chosen, as discussed in Section~\ref{sec:hPrimaryData}.
The concept of the swept rule can be applied in many ways with various design choices.

Beyond the ordering of computations, however, the swept scheme uses the same numerical method as the classic decomposition scheme. 
Briefly, for a one-dimensional domain, the heterogeneous one-dimensional swept rule begins by partitioning the computational grid and allocating space for the working array in each process. 
In this case, the working array is of type \texttt{states}, a C struct that contains the dependent and intermediate variables needed to continue the procedure from any time step. 

The working array size is determined by the number of dependent domains, or blocks, that a process controls ($N_{\text{blocks}}$) and the number of spatial points\slash threads within a domain ($N_{\text{threads}}$).\footnote{Here we use ``block'' and ``domain'' interchangeably to represent a domain of dependence; the term comes from the GPU\slash CUDA construct representing a collection of threads.}
The program allocates space for
$N_{\text{blocks}} \times N_{\text{threads}} + (N_{\text{threads}}+2) / 2$ spatial points and initializes the first $N_{\text{blocks}} \times N_{\text{threads}} + 2$ points.

The initialized points require two extra slots so the edge domains can build a full domain
width on their first step. 
Interior domains in the process share their edges with their neighbors; there is no risk of race conditions since even the simplest numerical scheme requires at least two values in the \texttt{state} struct, which allows the procedure to alternate reading and writing those values. 
Therefore, even as a domain writes an edge data point its neighbor must read, the value the neighbor requires is not modified.

The first cycle completes when each domain has progressed to the sub-time step
$N_{\text{threads}} / 2$, where it has computed two values at the center of the spatial domain.
At this point each process passes the first $N_{\text{threads}} / 2 + 1$ values in its array
to the left neighboring process. Each process receives the neighbor's buffer and places it in
the last $N_{\text{threads}} / 2 + 1$ slots; that is, starting at the $N_{\text{blocks}} \times N_{\text{threads}}$ index. It proceeds by performing the same computation on the centerpoints, starting at global index $N_{\text{threads}} - 1$ (adjusted index
$N_{\text{threads}} / 2 - 1$), of the new array and filling in the uncomputed grid points at
successive sub-time steps with a widening spatial window until it reaches a sub-time step that
has not been explored at any spatial point and proceeds with a contracting window.
Geometrically, the first cycle completes a triangle and the second cycle completes a diamond.
After completing the diamond, the program passes the last $N_{\text{threads}} / 2 + 1$ time steps in the array and inputs the received buffer starting at position 0. Now it performs the diamond procedure again, with identical global and adjusted indices starting at index $N_{\text{threads}} / 2 - 1$.

The procedure continues in this fashion until reaching the final time step, when it stops after the expanding window reaches the domain width and outputs the now-current solution at the same time step within and across all domains and processes. Therefore, the triangle functions are only used twice if no intermediate time step results are output, while the rest of the cycles are completed in a diamond shape.

Our program uses the MPI\allowbreak+CUDA paradigm and assigns one MPI process to each CPU core.
We considered using an MPI\allowbreak+OpenMP\allowbreak+CUDA paradigm by assigning an
MPI process to each socket and launching threads from each process to occupy the individual cores, but recent work has shown that this approach rarely improves performance on clusters of limited size for finite volume or finite difference solvers~\cite{IDAHO_MPI_CUDA,PerfAnalysisHetero}. 
This conclusion has led widely used libraries, such as PETSc, to opt against a paradigm of threading within processes~\cite{MillsPetsc}, and we followed this decision.

\subsection{Experimental method}
\label{sec:ExpMethod}

We will address the questions presented in Section~\ref{sec:obj1} by varying
three elements of the swept decomposition: block size (the number of spatial points in each domain), GPU work factor (ratio of the number of spatial points assigned to a GPU to those assigned to a single CPU process), and grid size. 
We repeatedly executed our two test equations, the heat equation and Euler equations, over the experimental domain of these variables using the swept and classic decomposition methods.

In our one-dimensional swept program for heterogeneous systems, \texttt{hSweep}, the size of
the domain-of-dependence or ``block'' is synonymous with number of threads per block, because it launches the solution of each domain using a block of threads on the GPU, where each thread handles one spatial point. 
In GPU\slash CUDA terms, a block is an abstract grouping of threads that share an execution setting, a streaming multiprocessor, and access to a shared-memory space, which is a portion of the GPU L1 cache. 
\texttt{hSweep} uses the swept rule to avoid communication between devices and processes and exploits the GPU memory hierarchy to operate on shared-memory quantities closer to the processor. 
Since this multi-level memory scheme influences the swept-rule performance and GPU execution, the resulting effects are difficult to predict.

The independent variables grid size and GPU work factor are more straightforward: the grid size is the total number of spatial points in the simulation domain, and the GPU work factor is the ratio of the computational grid assigned to the GPU to that assigned to the CPU:
\begin{equation}
    WF = \frac{N_{\text{GPU}}}{N_{\text{CPU}}} \;,
\end{equation}
where $N_{\text{GPU}}$ is the number of grid points assigned to the GPU and $N_{\text{CPU}}$ is the number assigned to a single CPU process\slash core.
We express the GPU work factor as the ratio of the number of domains-of-dependence assigned to the GPU to those assigned to a single MPI process (on a CPU core). 
Since a GPU can handle a larger portion of the total grid than a single MPI process, GPU work factor is specified as an integer greater than one.

In our previous study of the swept rule on a single GPU~\cite{OurJCP}, the properties of GPU architecture clearly defined the experimental domain. Here, because a warp contains 32 threads and a block cannot exceed 1024 threads, we constrained the number of threads per block to be a multiple of 32 from \numrange{32}{1024}; this is also the width of the domain of dependence. To enforce regularity, we constrained our experimental problem size---the number of spatial points in the grid---to be a power of 2 between \num{1024} and $2^{21}$.

The addition of GPU work factor as an independent variable further complicates the experimental domain. 
While our experiments are constrained by GPU architecture in threads per block and by the number of processes and blocks in problem size, we initially have no clear indication of what the experimental limits of GPU work factor should be.
We thus considered a wide range covering 0 to 100 in increments of 5.

In this study, we solve the one-dimensional heat equation using a first-order forward in time, central in space method, and the Euler equations using a second-order finite-volume scheme with minmod limiter. 
Explanations of these methods can be found in the appendix of our previous study~\cite{OurJCP}.

\subsection{Primary data structure experiment}
\label{sec:hPrimaryData}

Implementing the swept rule for problems amenable to single-step PDE schemes is straightforward, but dealing with more realistic problems often requires more complex, multi-step numerical schemes. 
Managing the working array and developing a common interface for these schemes provokes design decisions that substantially impact performance. 
In this article we consider two strategies for dealing with this complexity: ``\texttt{flattening}'' and ``\texttt{lengthening}'', distinguished via code snippets in Figure~\ref{fig:flatlong}.

\begin{figure}[hbt]
  \centering
  \begin{subfigure}[t]{0.46\textwidth}
    \begin{lstlisting}[language=C++, gobble=6]
      __global__ void classicStep(const double *s_in, double *s_out, bool final) {
        int gid = blockDim.x * blockIdx.x + threadIdx.x;
        // number of spatial points - 1
        int lastidx = ((blockDim.x * gridDim.x));
        int gids[5];

        for (int k = -2; k < 3; k++) {
          gids[k+2] = (gid + k) % lastidx;
        }

        // Final is false for predictor step, true otherwise.
        if (final) {
          s_out[gid] += finalStep(s_in, gids);
        } else {
          s_out[gid]  = predictorStep(s_in, gids);
        }
      }
    \end{lstlisting}
    \caption{\texttt{flattening} method. The sub-timesteps are compressed to a step with a
		wider stencil. The two arrays which alternate reading and writing are explicitly passed
		and traded in the calling function.}
    \label{fig:flatPseudo}
  \end{subfigure}
  ~~
  \begin{subfigure}[t]{0.46\textwidth}
    \begin{lstlisting}[language=C++, gobble=6]
      // Q = {rho, rho*u, rho*E}
      struct states {
        double3 Q[2]; // State Variables
        double Pr; // Pressure ratio
      };

      __device__ __host__
      void stepUpdate(states *state, const int idx, const int tstep) {
        int ts = tstep % 4; // 4 is number of steps in cycle
        if (tstep & 1)  pressureRatio(state, idx, ts);
        else            eulerStep(state, idx, ts);
      }

      __global__ void classicStep(states *state, const int tstep) {
        int gid = blockDim.x * blockIdx.x + threadIdx.x + 1;
        stepUpdate(state, gid, tstep);
      }
    \end{lstlisting}
  \caption{\texttt{lengthening} method. The \texttt{states} struct contains all
	information to step forward at any point. A user only needs to write the
	\texttt{eulerStep()} and \texttt{pressureRatio()} functions and accessing the correct members
	based on the time step.}
  \label{fig:Longpseudo}
 \end{subfigure}
 \caption{Brief code skeletons for the \texttt{flattening} and \texttt{lengthening} methods,
 applied to solving the Euler equations using classic domain decomposition.}
 \label{fig:flatlong}
\end{figure}

The \texttt{flattening} scheme flattens the domain of dependence in the time dimension by using wider stencils and fewer sub-timesteps. 
This strategy is more memory efficient for the working array, which contains instances of the primary data structure at each spatial point, but it cannot easily accommodate different methods and equations. 
It also introduces additional complexity from parsing the arrays and requires additional register memory for function and kernel arguments and ancillary variables. 
Figure~\ref{fig:flatPseudo} depicts the \texttt{flattening} approach applied to the Euler equations using classic domain decomposition.

In the new implementation shown here we use the \texttt{lengthening} strategy, also referred to as ``atomic decomposition'', which is instantiated as a struct to generalize the stages into a user-defined data type~\cite{WangDecomp} as shown in Figure~\ref{fig:Longpseudo}.
It requires more memory for the primary data structure; for instance, our \texttt{flattening} version of the Euler equations carries six doubles per spatial point since the pressure ratio used by the limiter was rolled into the flattened step. 
By restricting the stencil to three points, the \texttt{lengthening} method requires the pressure ratio to be stored and passed through the memory hierarchy, meaning the data structure carries seven doubles per spatial point for the Euler equations.

To gauge the influence of our choice for primary data structure, we implemented each combination of the classic and swept decomposition techniques using the \texttt{flattening} and \texttt{lengthening} data structures. 
We applied these methods to solve the Euler equations using the discretization and conditions described in Section~\ref{sec:ExpMethod}.
For these tests we used a workstation with an \WCPU{} and a single \WGPU{} GPU. 
(This differs from the other results shown in this article, which use CPUs and GPUs in concert on a heterogeneous platform.)

\begin{figure}[htbp]
\centering
 \begin{subfigure}[t]{.8\textwidth}
  \centering
  \includegraphics[width=\textwidth]{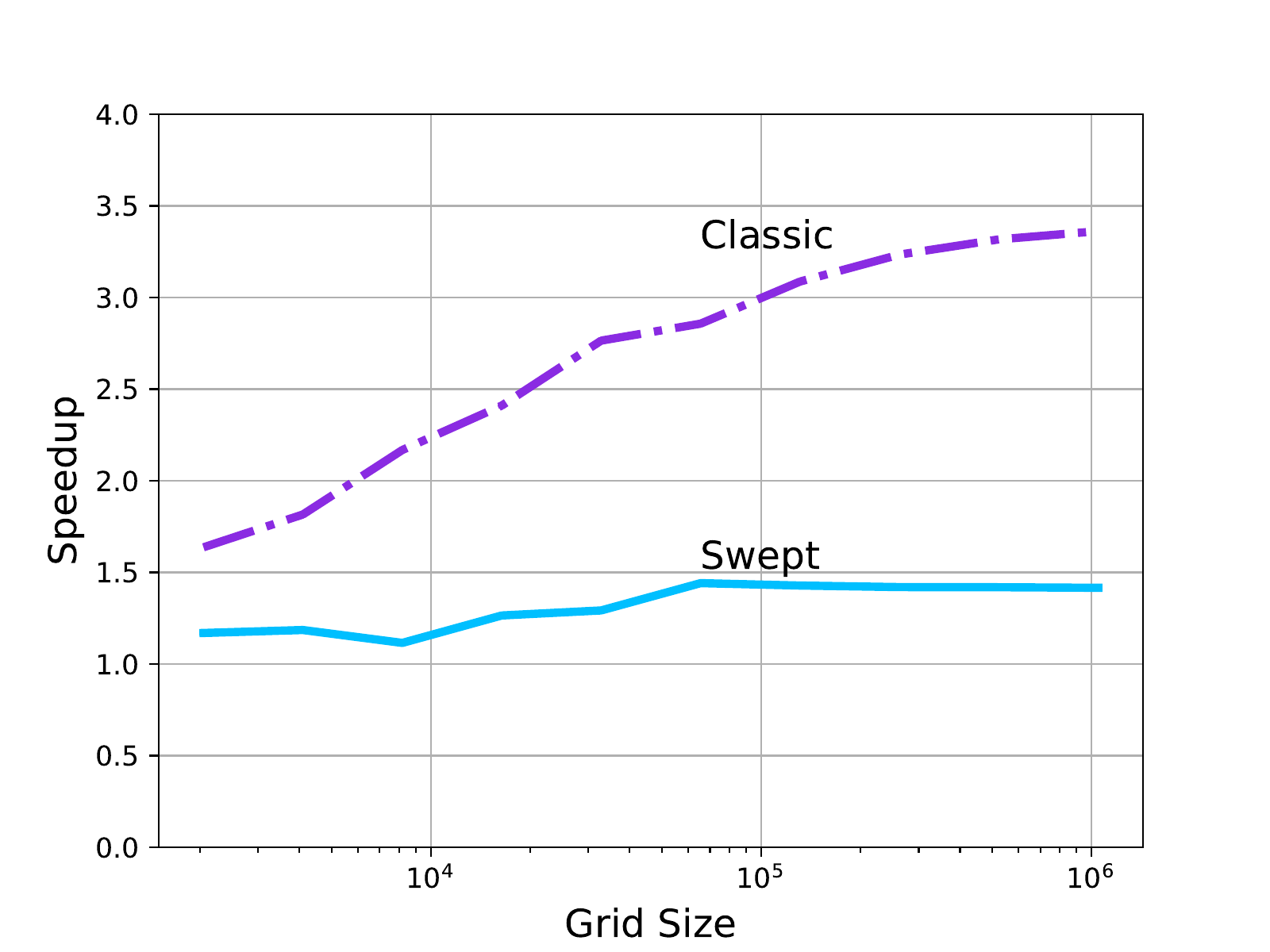}
   \caption{Speedup of the \texttt{flattening} method vs. \texttt{lengthening} method for the \texttt{Swept} and \texttt{Classic} schemes.}
  \label{fig:flatLongSpeedup}
 \end{subfigure}
 \\
  \begin{subfigure}[t]{.8\textwidth}
  \centering
  \includegraphics[width=\textwidth]{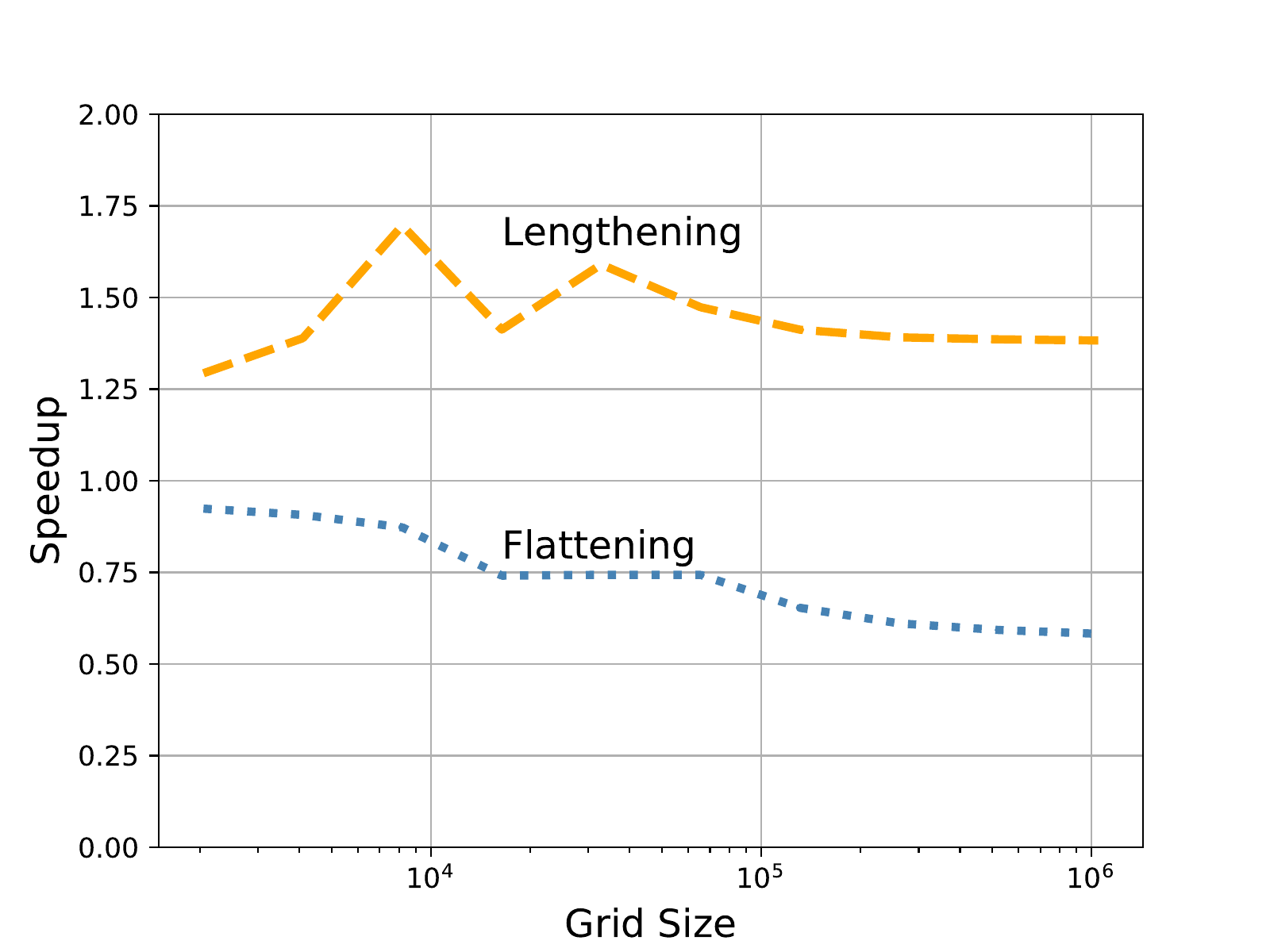}
  \caption{Speedup of the \texttt{Swept} scheme vs. \texttt{Classic} for the \texttt{flattening} and \texttt{lengthening} methods.}
  \label{fig:sweptClassicSpeedup}
 \end{subfigure}
\caption{Performance comparison of solving the Euler equations using the \texttt{flattening} and \texttt{lengthening} methods with the \texttt{Swept} and \texttt{Classic} schemes.}
\label{fig:data-structure}
\end{figure}

Figure~\ref{fig:data-structure} compares the performance of the data-structure experiments solving the Euler equations, with all computations performed using a single GPU.
Figure~\ref{fig:flatLongSpeedup} shows the speedup of the \texttt{flattening} method against the \texttt{lengthening} method for each decomposition scheme:
\begin{equation}
    S_{\text{flat}} = \frac{\text{time}_{\text{length}, s}}{\text{time}_{\text{flat}, s}} \;,
\end{equation}
where $s$ is the decomposition scheme (\texttt{Classic} or \texttt{Swept}).
Figure~\ref{fig:sweptClassicSpeedup} shows the speedup of the \texttt{Swept} scheme against the \texttt{Classic} scheme for each method, calculated as 
\begin{equation}
    S_{\text{swept}} = \frac{\text{time}_{m, \text{classic}}}{\text{time}_{m, \text{swept}}} \;,
\end{equation}
where $m$ is the method (\texttt{lengthening} or \texttt{flattening}).
These plots use the best run times for a given grid size over \numlist{64;128;256;512;1024} threads per block. 

Figure~\ref{fig:flatLongSpeedup} shows that the \texttt{flattening} strategy makes both decomposition schemes perform slightly faster,
and the benefit increases with number of spatial points.
The method improves the \texttt{Classic} scheme more, raising the speedup from about \SI{1.6}{$\times$} to \SI{3.4}{$\times$} with growing grid size, while the \texttt{Swept} scheme with \texttt{flattening} improves from about \SI{1.2}{$\times$} to \SI{1.5}{$\times$} faster.
The \texttt{lengthening} method performs worse for both decomposition schemes due to the GPU architecture: the array of structures used in \texttt{lengthening} amplifies the performance sensitivity to irregularity in memory-access patterns.
As a result, we cannot easily generalize these results to the heterogeneous systems of primary interest here, though clearly the method choice impacts GPU performance with both decomposition schemes.

The extra memory requirements of the \texttt{lengthening} method also consume limited shared-memory resources on the GPU, which diminishes both occupancy (the number of threads active on the GPU at any given time) for \texttt{Swept} and the L1 cache capacity used to accelerate global-memory accesses on Kepler-generation GPUs for \texttt{Classic}. 
While locality is a significant issue for effective CPU memory accesses, it impacts GPU performance more.

These issues explain the general benefit of the \texttt{flattening} strategy, but they do not explain why these benefits are more pronounced for \texttt{Classic}. First, the \texttt{lengthening} strategy requires more sub-steps per time step to limit the stencil to three points.
This does not affect the number of kernels launches for \texttt{Swept}, but may increase the occurrence of these events for \texttt{Classic}.
For the Euler equations, our scheme uses four sub-steps per time step using \texttt{lengthening} and two sub-steps per time step using \texttt{flattening}.
This causes the \texttt{Classic lengthening} scheme to launch twice as many kernels as it would with \texttt{flattening}.
Also, the aforementioned array-structure paradigm used in the \texttt{lengthening} strategy increases the stride for memory accesses.
This has little effect on the \texttt{Swept} scheme because it uses mostly shared memory. 
In contrast, though this access pattern does not produce bank conflicts, it does prevent global memory accesses from coalescing, incurring a significant cost in the \texttt{Classic} program.

Figure~\ref{fig:sweptClassicSpeedup} shows how the choice of primary data structure affects the speedup of \texttt{Swept} compared with \texttt{Classic} for solving the Euler equations. 
These results match what we found in our previous study~\cite{OurJCP}: \texttt{Swept} performs worse than \texttt{Classic} using the \texttt{flattening} method on a single GPU solving the Euler equations. 
However, examining the current results, the swept decomposition scheme improves performance over the classic approach when using the \texttt{lengthening} strategy.

In any experimental algorithm, especially those involving emerging, parallel computational platforms, performance depends on a multitude of implementation details, and we feel that it is important to present these findings so that others who implement the swept rule will have a more thorough understanding of the tradeoffs inherent in particular design choices.
For this study, we selected the \texttt{lengthening} method due to its benefits for the swept scheme, as well as better extensibility and regularity.
However, when discussing the overall results, we need to keep in mind that the \texttt{Classic} scheme could perform 1.5--3.5 faster on a GPU if using the \texttt{flattening} method.

\section{Results}
\label{sec:hResults}


We compiled all test programs with the CUDA compiler \texttt{nvcc v8} and launched using
\texttt{mpirun v3.1.4}.
Our study collects the average time per time step over \num{6000} time steps. 
The timing measurements include the onetime costs of device memory allocation,
plus initial and final host-device memory transfers; 
this does not include the time cost of
host-side memory management, grid generation, file I/O, and initial condition calculations.
The heterogeneous swept rule algorithms and test cases were implemented in
\texttt{hSweep v2.0}~\cite{hSweepz}.

We performed the tests on the \texttt{Classic} and \texttt{Swept}
programs on two nodes of the Oregon State University College of Engineering cluster. 
Each node contains two sockets with \CCPU{} 
ten-core processors each operating at \SI{2.60}{\giga\hertz}. 
An~\CGPU{} GPGPU is available to one of these nodes through a PCI connection.

\subsection{Effects of launch conditions}

\begin{figure}[htbp]
 \centering
 \begin{subfigure}[t]{.8\textwidth}
  \centering
  \includegraphics[width=\textwidth]{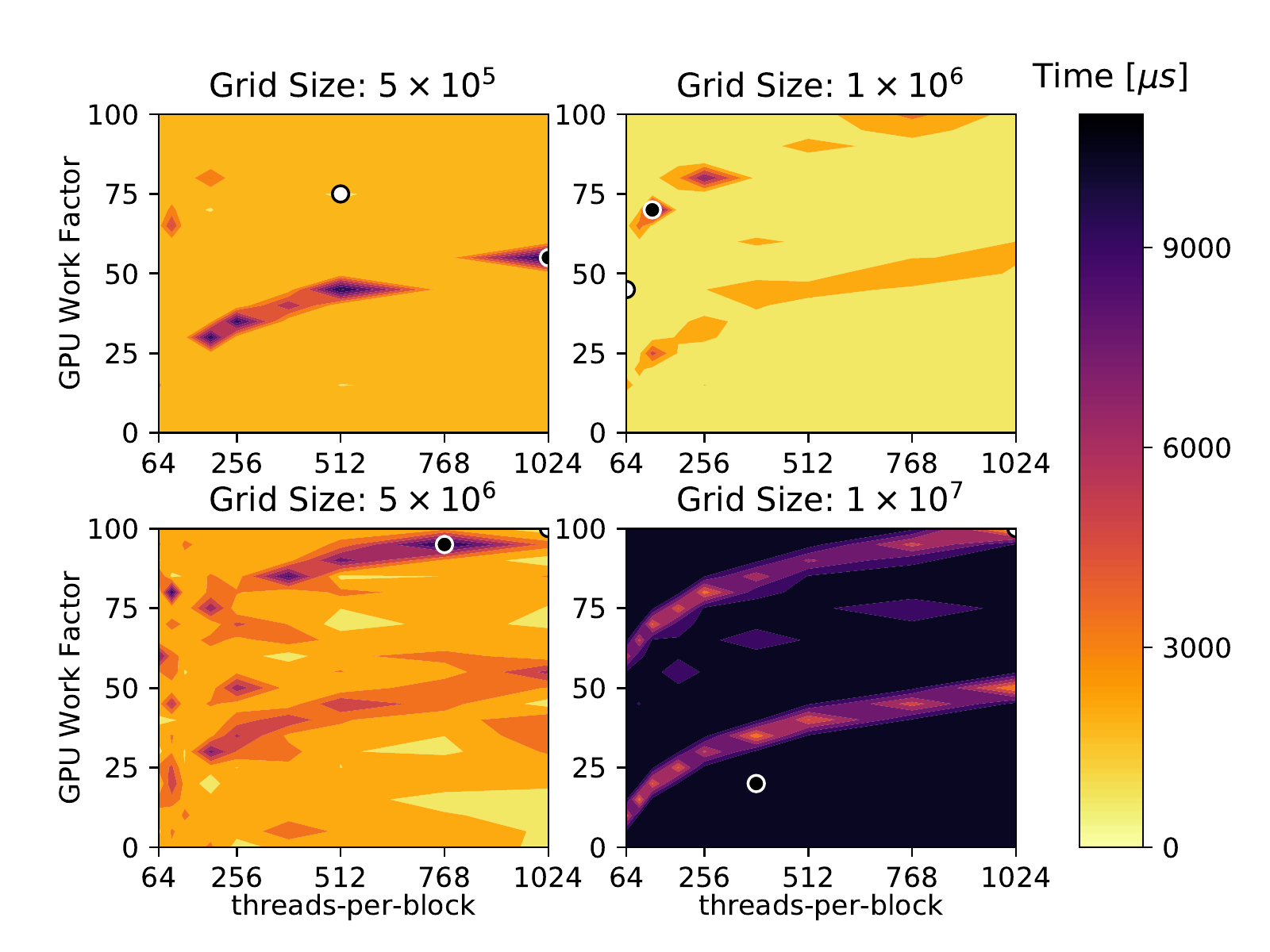}
  \caption{Classic decomposition}
  \label{fig:HeatContourC}
 \end{subfigure}
 \\
 \begin{subfigure}[t]{.8\textwidth}
  \centering
  \includegraphics[width=\textwidth]{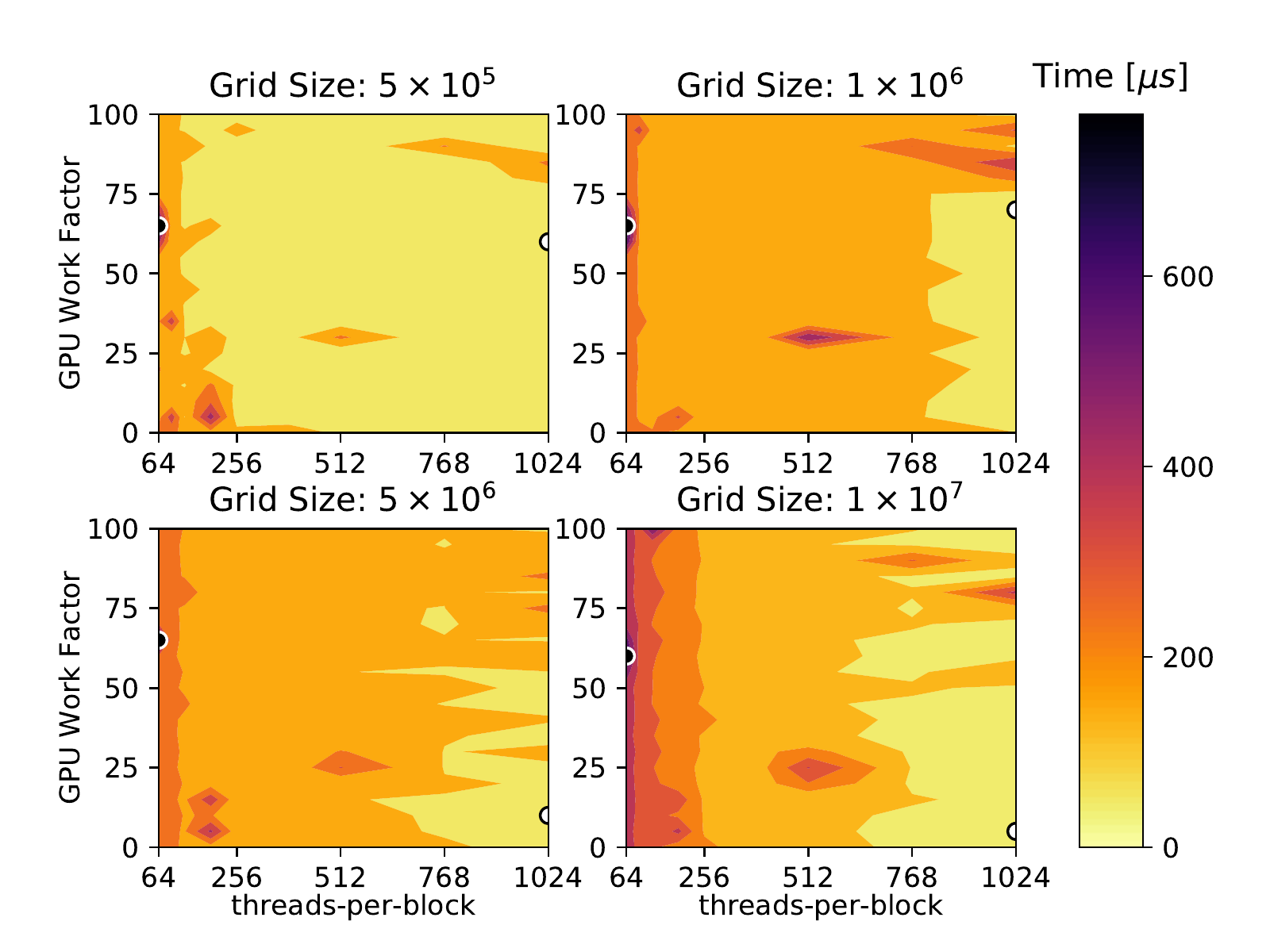}
  \caption{Swept decomposition}
  \label{fig:HeatContourS}
 \end{subfigure}
 \caption{Computational time per time step for the heat equation at four grid sizes, varying GPU work factor and threads per block (lower is faster): \numlist{5e5;1e6;5e6;1e7}.
 The white dot signifies the best performance and the black dot the worst performance.}
 \label{fig:HeatContoursFull}
\end{figure}

\begin{figure}[htbp]
 \centering
 \begin{subfigure}[t]{.8\textwidth}
  \centering
  \includegraphics[width=\textwidth]{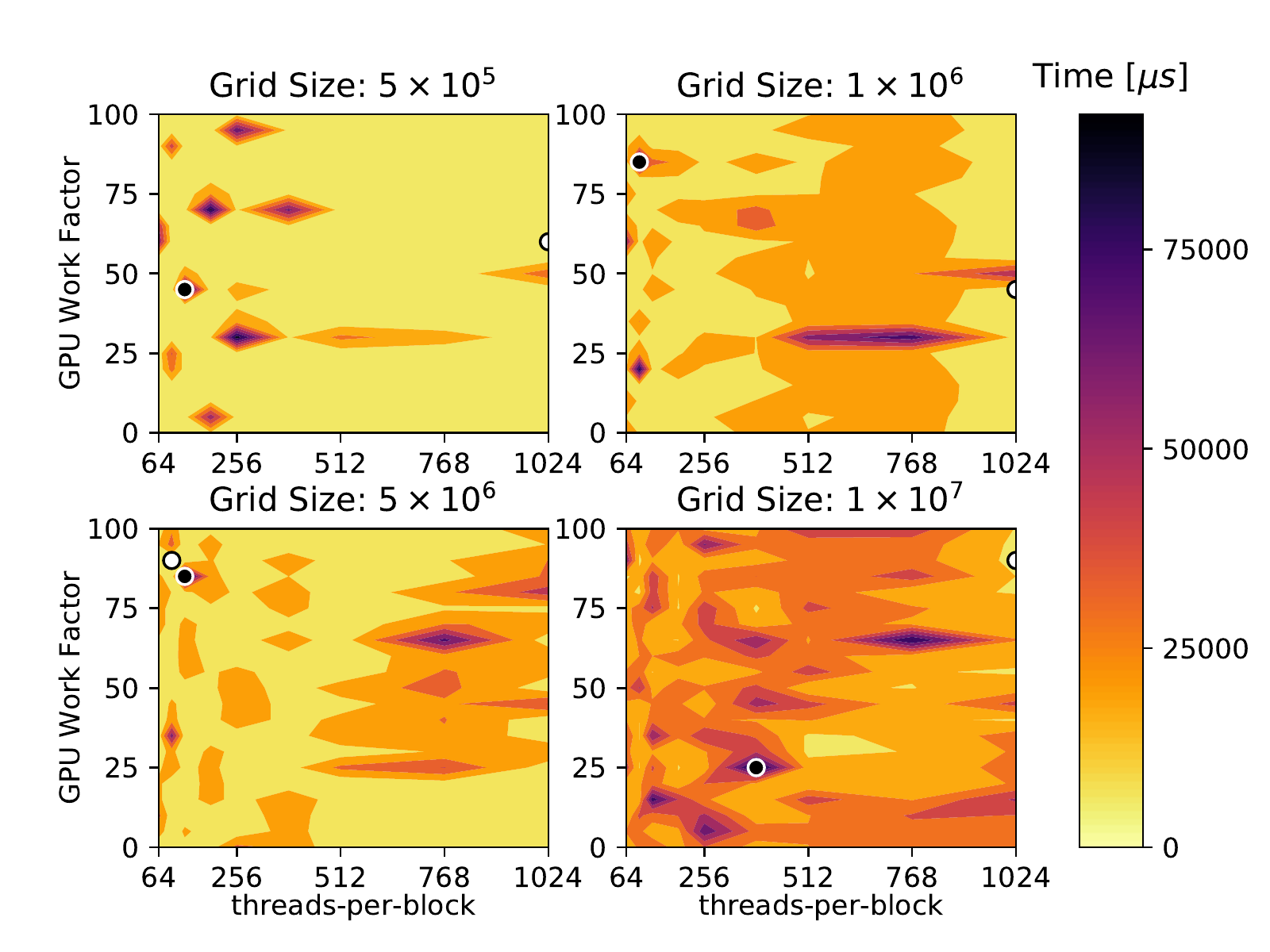}
  \caption{Classic decomposition}
  \label{fig:EulerContourC}
 \end{subfigure}
 \\
 \begin{subfigure}[t]{.8\textwidth}
  \centering
  \includegraphics[width=\textwidth]{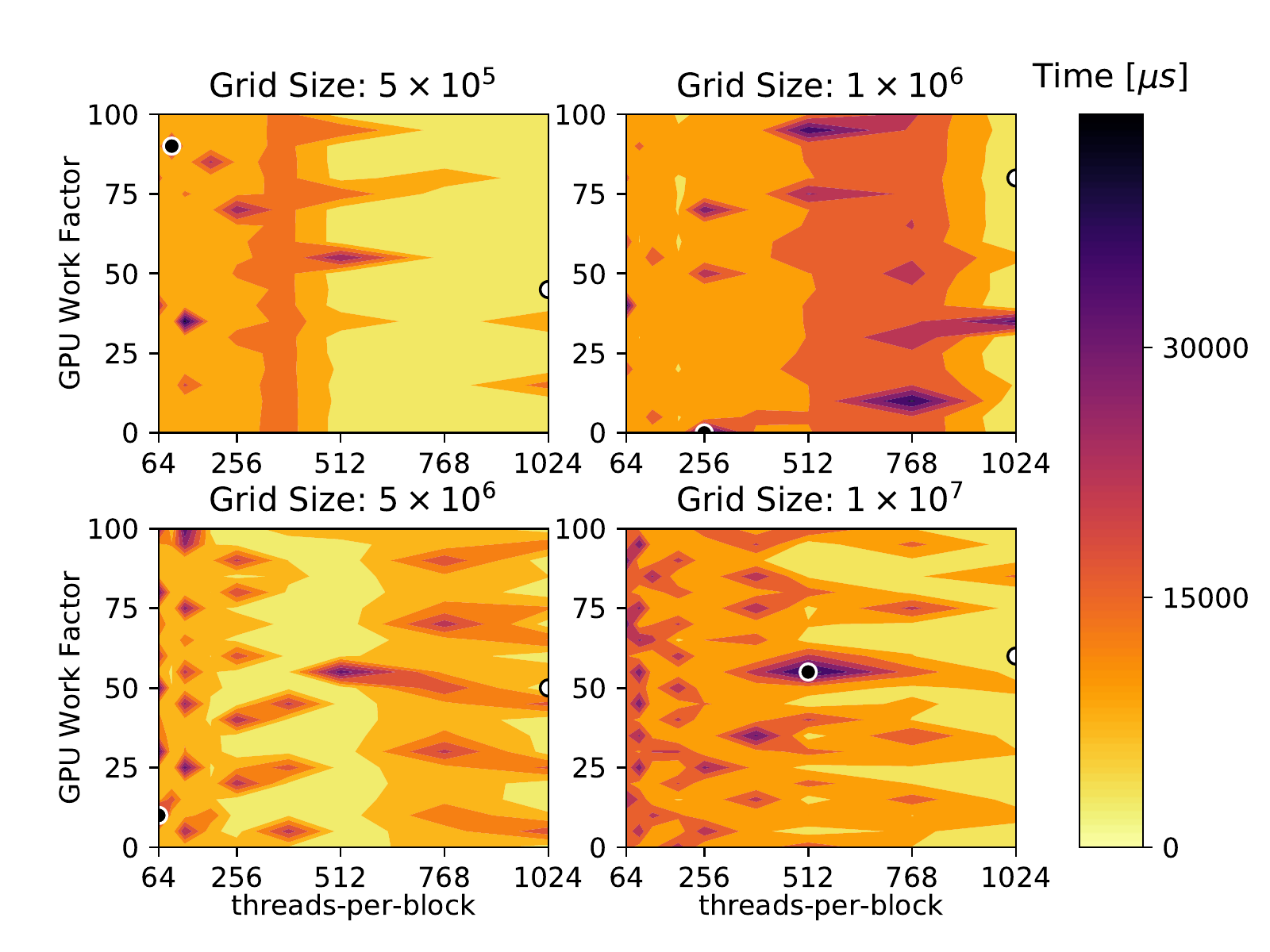}
  \caption{Swept decomposition}
  \label{fig:EulerContourS}
 \end{subfigure}
 \caption{Computational time per time step for the Euler equations at four grid sizes, varying GPU work factor and threads per block (lower is faster): \numlist{5e5;1e6;5e6;1e7}.
 The white dot signifies the best performance and the black dot the worst performance.}
 \label{fig:EulerContoursFull}
\end{figure}

\begin{figure}[htbp]
 \centering
 \includegraphics[width=0.9\textwidth]{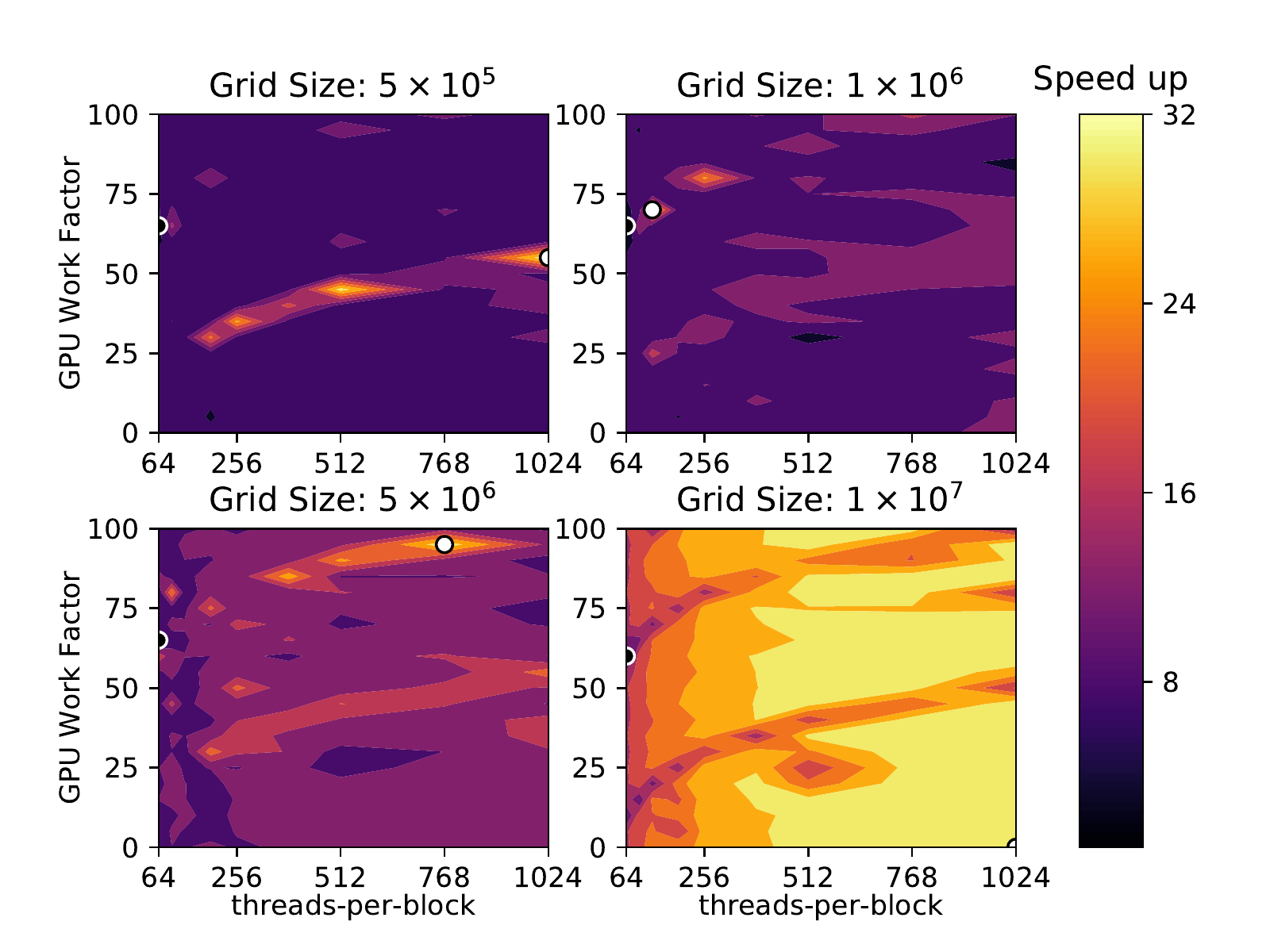}
 \caption{Speedup of \texttt{Swept} to \texttt{Classic} decomposition (i.e., runtime of \texttt{Classic} / runtime of \texttt{Swept}) for the heat equation at four grid sizes, varying GPU work factor and threads per block (higher means faster speedup of \texttt{Swept} decomposition): \numlist{5e5;1e6;5e6;1e7}.
 The white dot signifies the best performance and the black dot the worst performance.}
 \label{fig:HeatContoursSpeedup}
\end{figure}

\begin{figure}[htbp]
 \centering
 \includegraphics[width=0.9\textwidth]{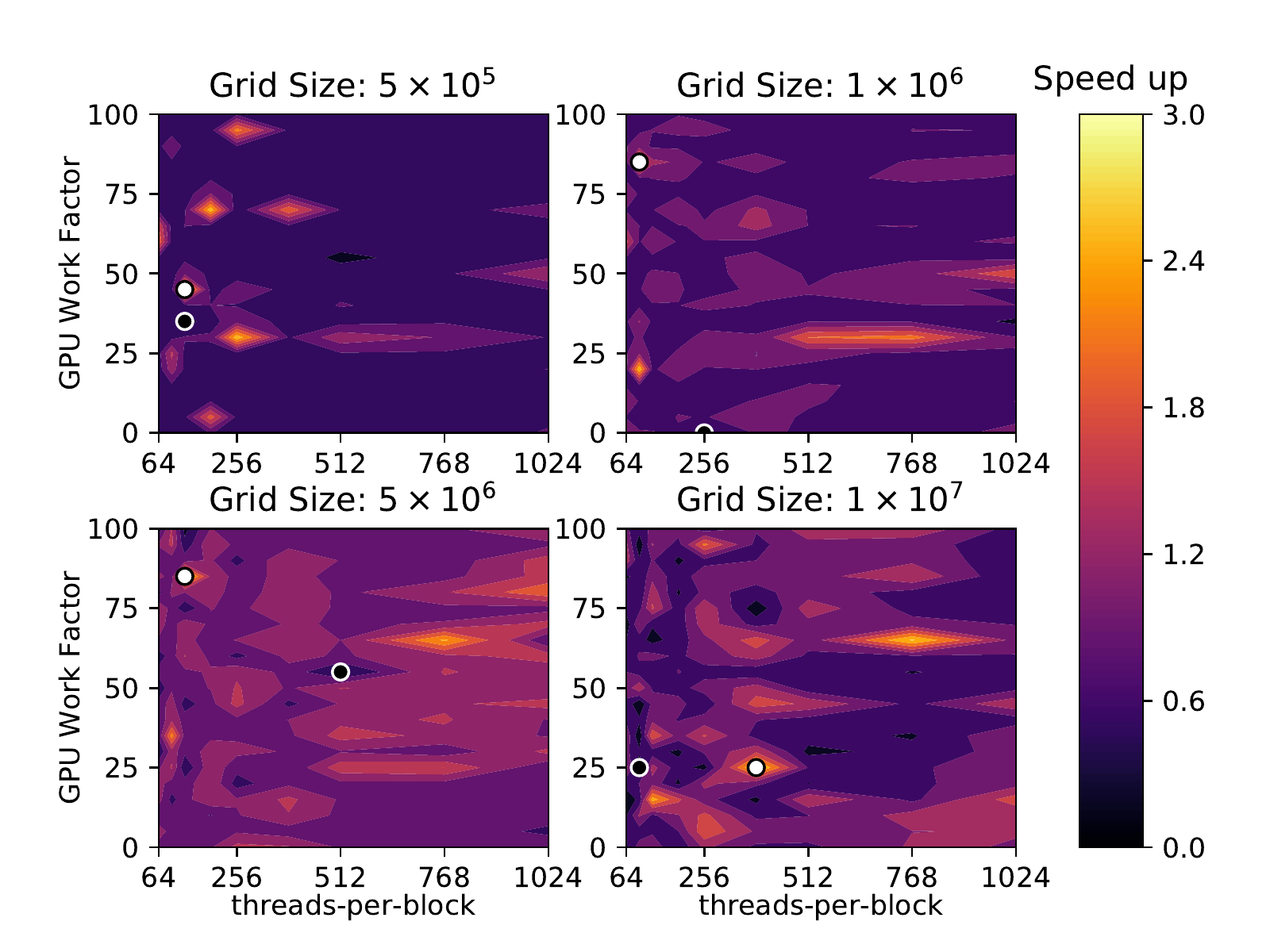}
 \caption{Speedup of \texttt{Swept} to \texttt{Classic} decomposition (i.e., runtime of \texttt{Classic} / runtime of \texttt{Swept}) for the Euler equations at four grid sizes, varying GPU work factor and threads per block (higher means faster speedup of \texttt{Swept} decomposition): \numlist{5e5;1e6;5e6;1e7}.
 The white dot signifies the best performance and the black dot the worst performance.}
 \label{fig:EulerContoursSpeedup}
\end{figure}

First, we measured the performance of the \texttt{Swept} and \texttt{Classic} decomposition
schemes using the one-dimensional heat and Euler equations over the complete range of
operating conditions.
Figures~\ref{fig:HeatContoursFull} and \ref{fig:EulerContoursFull} show the average runtime per time step of the heat and Euler equations, respectively;
Figures~\ref{fig:HeatContoursSpeedup} and \ref{fig:EulerContoursSpeedup} show the speedup of the \texttt{Swept} approach for each problem.
The range of conditions covers a GPU work factor of 0 to 100, 64--1024 threads per block (limited by the hardware), and grid sizes of \numlist{5e5;1e6;5e6;1e7} points.
Notably, these results show that, particularly for the Euler equations, the launch conditions play an important role in the relative performance of the decomposition schemes.
The \texttt{Swept} scheme achieves best speedup for higher GPU work factors (about 50--95) for grid sizes of \numrange{5e5}{5e6}, but then smaller GPU work factors for for \num{1e6} points.
This may suggest that the largest number of grid points may be too many for a single GPU; however, locations of high speedup also appear for various combinations of GPU work factor and threads per block.
One consistent trend is lower speedup for the Euler equations with more threads per block (i.e., more points in the domain of dependence) at smaller grid sizes.
This behavior does not appear for solving the heat equation; in fact, higher threads per block generally lead to higher speedup for that problem.

For the heat equation, the performance of the \texttt{Swept} decomposition scheme is fairly insensitive to GPU work factor, but with better performance for more threads per block.
In contrast, \texttt{Classic} decomposition is more sensitive to combination of GPU work factor and number of threads per block; 
the speedup also reflects this sensitivity.
For the Euler equations, both decomposition schemes depend on the launch conditions in a complicated way, with combinations of both high and low performance evident.
However, for the Euler equations, the \texttt{Swept} scheme performs best with more threads per block.
Interestingly, the \texttt{Swept} scheme appears less sensitive to GPU work factor for both problems and all grid sizes.

\subsection{Comparison at best conditions}

\begin{figure}[htbp]
\centering
\begin{subfigure}[t]{0.48\textwidth}
    \includegraphics[width=\textwidth]{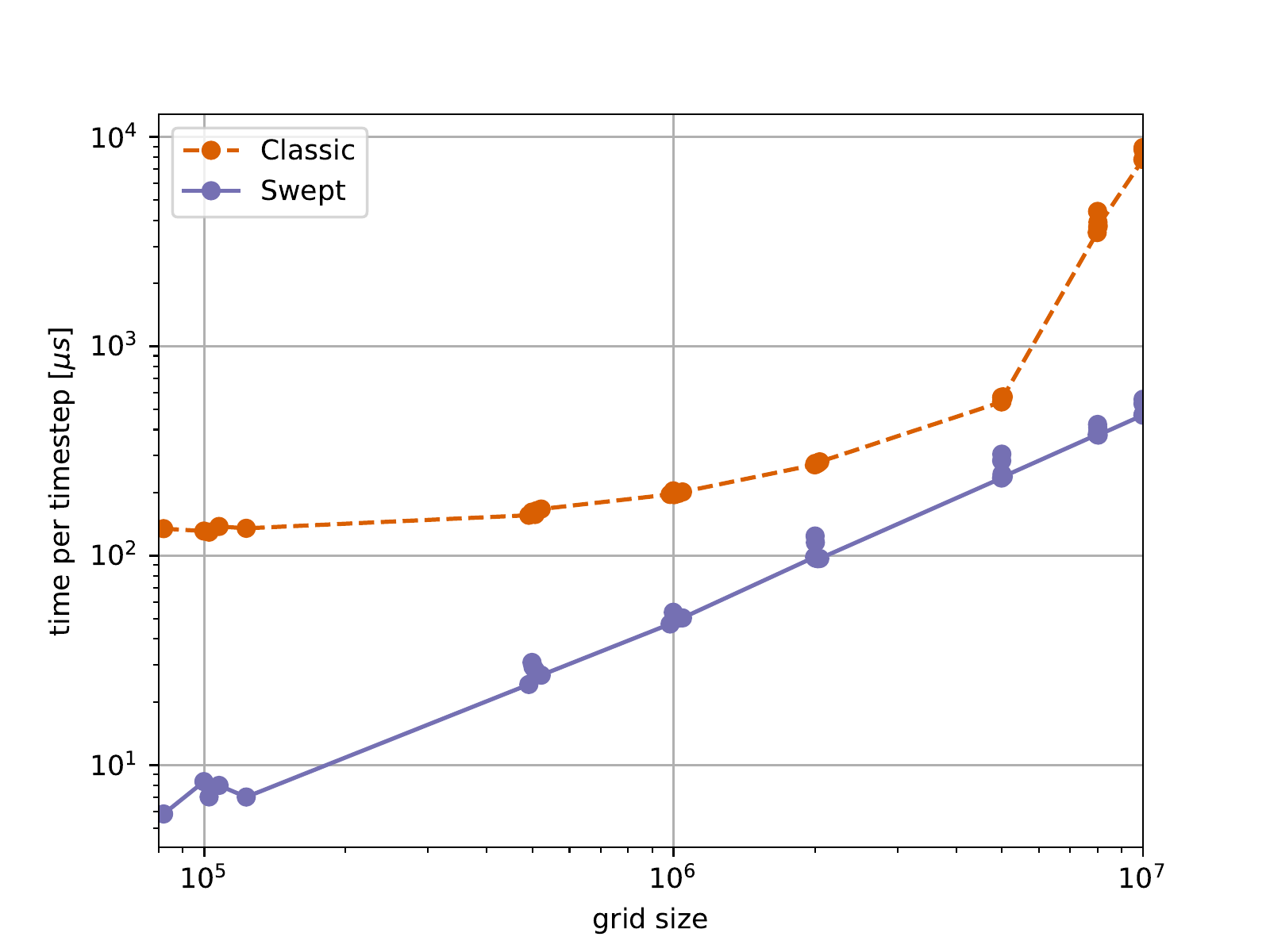}
    \caption{Time cost per time step of solving the heat equation using the best launch conditions.}
    \label{fig:BestRuntimeHeat}
\end{subfigure}
~
\begin{subfigure}[t]{0.48\textwidth}
    \includegraphics[width=\textwidth]{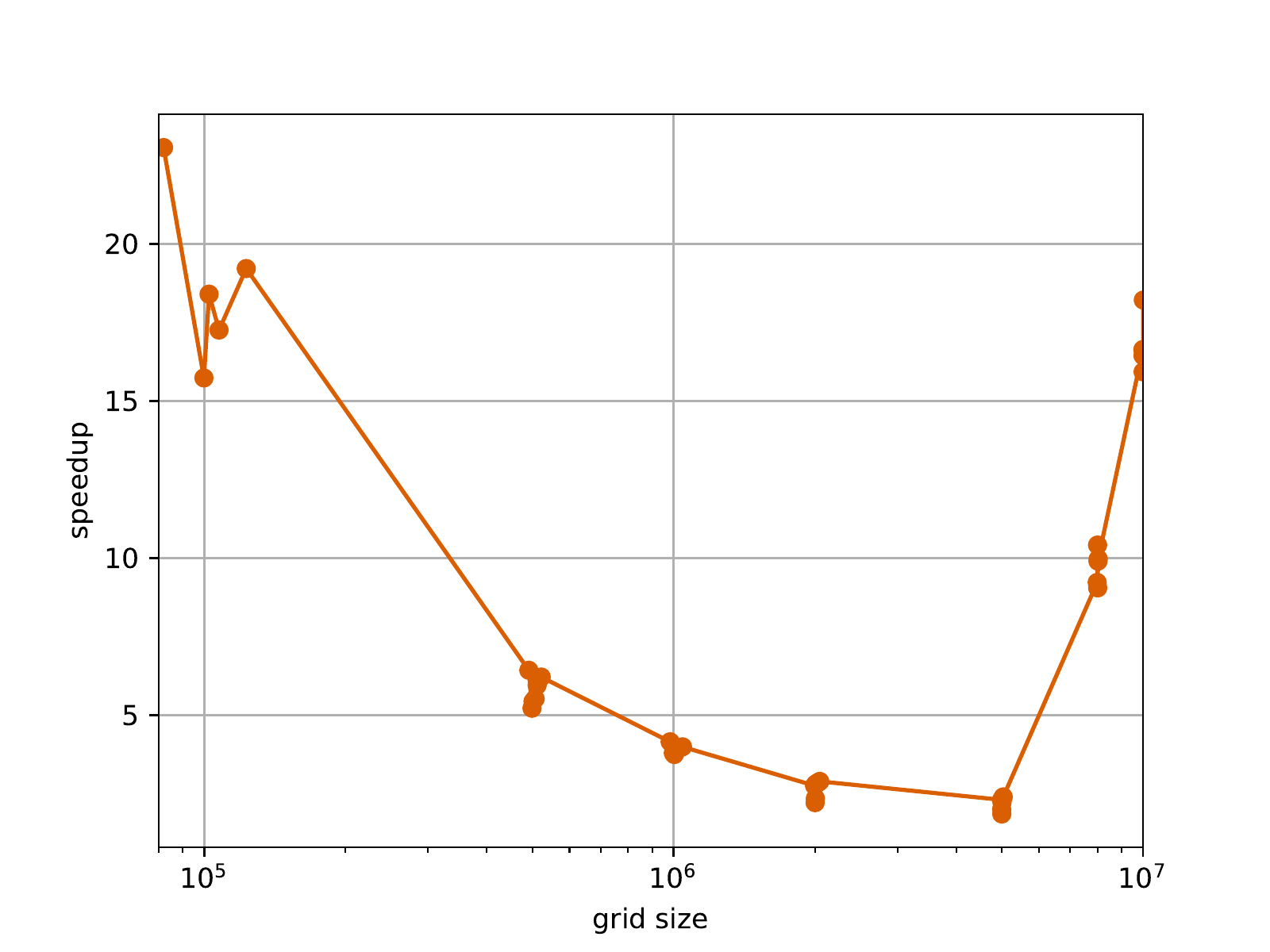}
    \caption{Speedup of \texttt{Swept} at best launch conditions.}
    \label{fig:BestSpeedupHeat}
\end{subfigure}
\caption{Performance comparison of \texttt{Swept} and \texttt{Classic} decomposition
methods solving the one-dimensional transient heat equation.}
\label{fig:HeatBest}
\end{figure}

\begin{figure}[htbp]
\centering
\begin{subfigure}[t]{0.48\textwidth}
    \includegraphics[width=\textwidth]{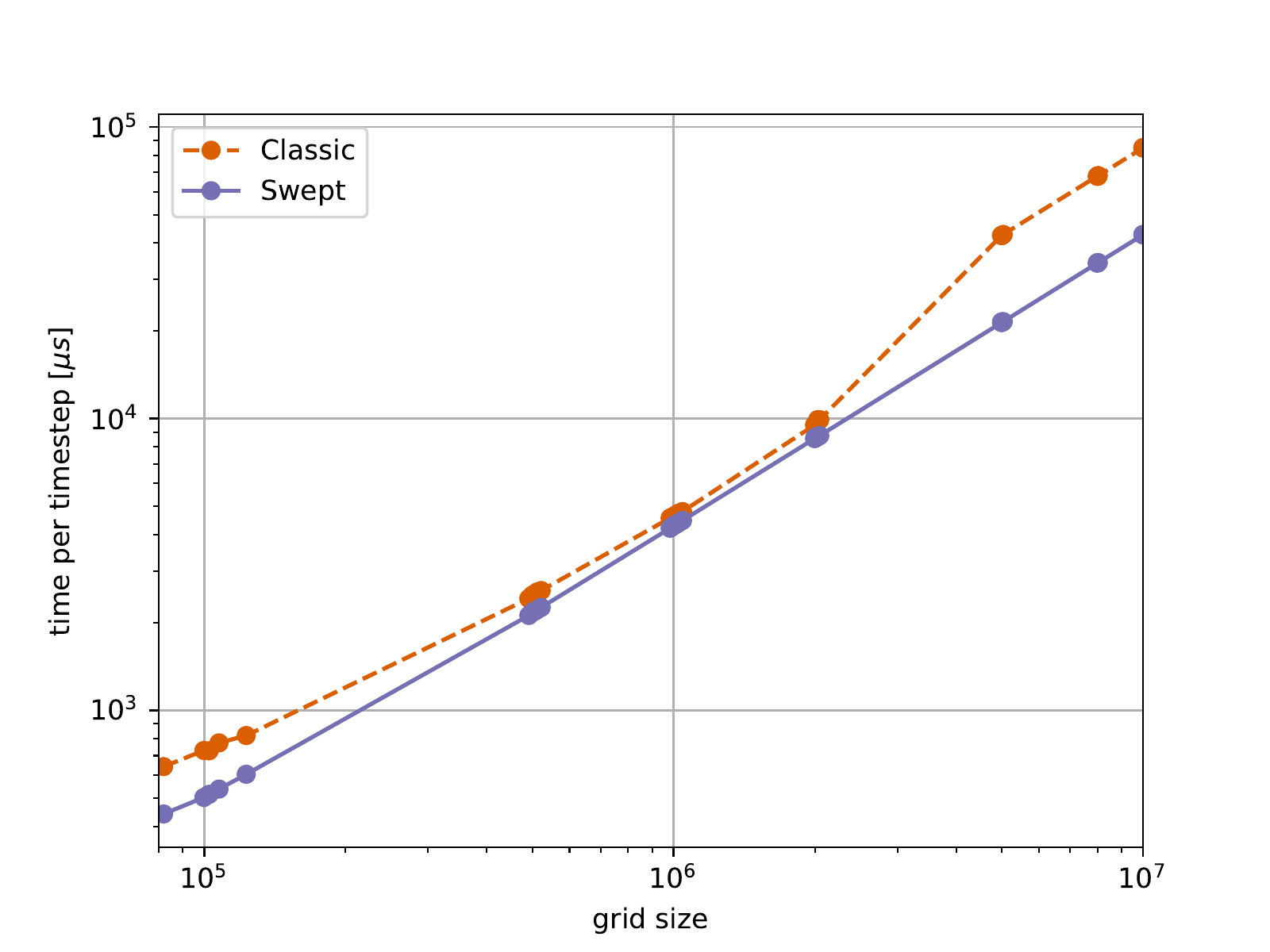}
    \caption{Time cost per time step using the best launch conditions.}
    \label{fig:BestRuntimeEuler}
\end{subfigure}
~
\begin{subfigure}[t]{0.48\textwidth}
    \includegraphics[width=\textwidth]{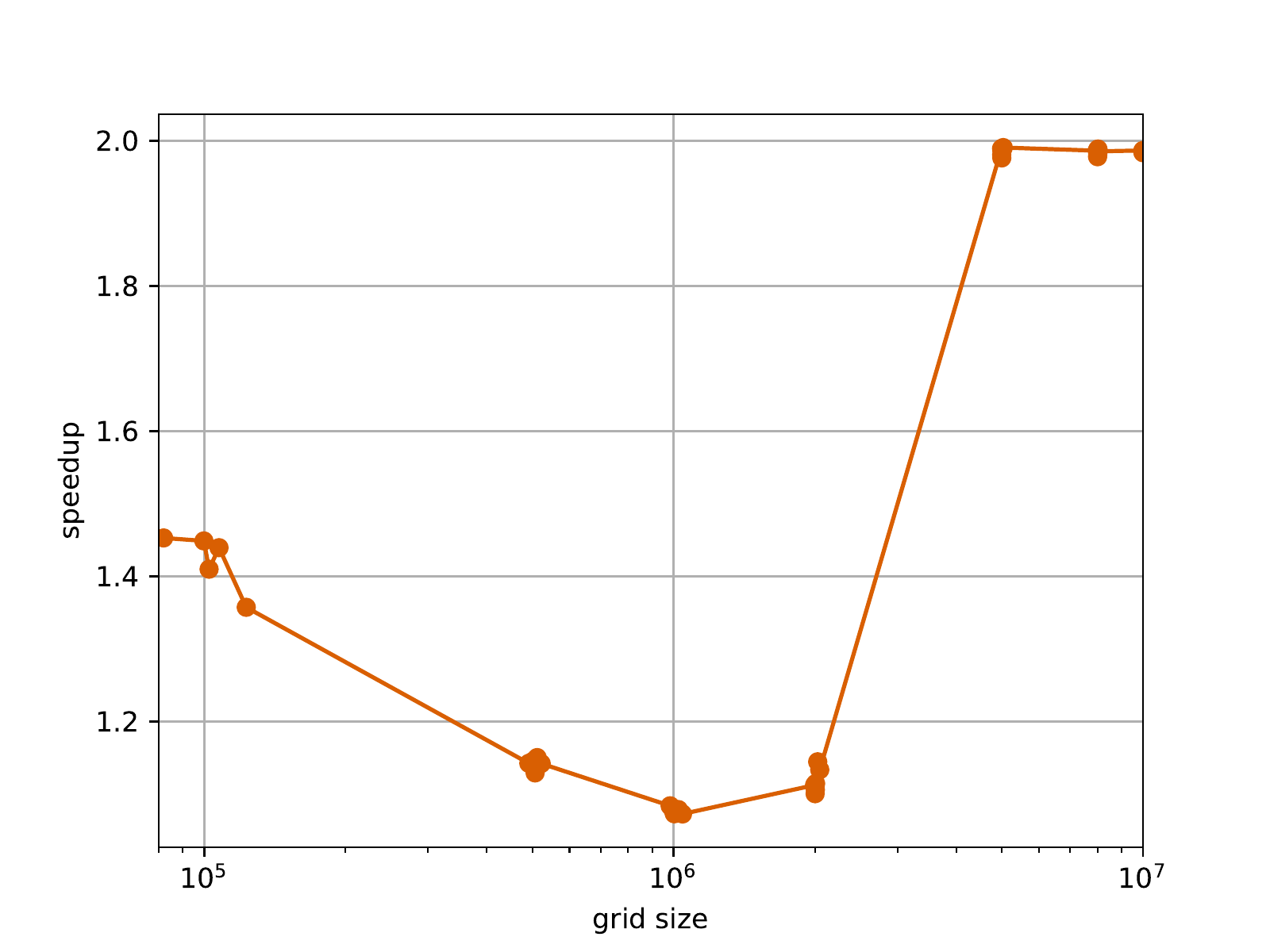}
    \caption{Speedup of \texttt{Swept} at best launch conditions.}
    \label{fig:BestSpeedupEuler}
\end{subfigure}
\caption{Performance comparison of \texttt{Swept} and \texttt{Classic} decomposition
methods solving the one-dimensional Euler equations.}
\label{fig:EulerBest}
\end{figure}

Next, we extracted the results for the best launch conditions at each grid size for the \texttt{Swept} and \texttt{Classic}.
Figures~\ref{fig:HeatBest} and \ref{fig:EulerBest} compare the computational time per time step of each scheme and the speedup of \texttt{Swept} for the heat equation and Euler equations, respectively.
For the one-dimensional heat equation, the \texttt{Swept} scheme achieves a speedup of between 1.9 and 23, depending on the grid size; the highest speedups appear for the smallest and largest grid sizes.
For the Euler equations, the \texttt{Swept} scheme offers a speedup of between about 1.1 and 2.0, with the best improvement at the largest grid sizes.
In both cases, the variations in speedup stem from the variable performance of the \texttt{Classic} decomposition.
In contrast, the \texttt{Swept} decomposition's performance closely follows a
power law for both problems, resulting from the fairly regular performance shown earlier.
In fact, we performed a least-squares fit of these results to a power law, as shown in Table~\ref{tab:tablefit}, and found excellent agreement ($R^2 > 0.99$ in both cases).

\begin{table}[htbp]
 \caption{Coefficients for power-law fit of computational time per time step in \si{\micro\second} as a function of grid size ($y = Ax^b$) for
 \texttt{Swept} performance at best launch configurations.}
 \label{tab:tablefit}
 \begin{center}
  \begin{tabular}{@{}l c c c@{}}
   \toprule
   Equation & $A$ (\si{\micro\second}) & $b$ & $R^2$ \\
   \midrule
   Heat  & \num{1.33e-4} & \num{0.937} & \num{0.997} \\
   Euler & \num{6.77e-3} & \num{0.970} & \num{0.999} \\
   \bottomrule
  \end{tabular}
 \end{center}
\end{table}

First, we discuss the results for the heat equation shown in Figure~\ref{fig:BestRuntimeHeat}.
We observed in previous studies~\cite{OurJCP,alhubail:16jcp} that \texttt{Swept} performs better than \texttt{Classic} decomposition when applied to simpler problems (i.e., the heat equation), but reaches maximum performance at smaller grid sizes. This difference in the capacity for increasing grid sizes narrows the performance advantage of \texttt{Swept} swiftly, and, since both algorithms scale similarly with grid size, the relative speedup of \texttt{Swept} declines as well.
However, \texttt{Swept} performs faster than \texttt{Classic} for all grid sizes.
We also see here that the \texttt{Classic} scheme becomes significantly slower at the largest grid sizes considered, while the \texttt{Swept} scheme continues its regular cost increase.  

Second, although the minimum grid size in this study is about \tx{100} larger
than our previous study~\cite{OurJCP}, and the maximum grid size is about \tx{10} larger, Figure~\ref{fig:BestSpeedupHeat} shows a similar trend in the speedup of \texttt{Swept} decomposition for the heat equation.
The heterogeneous \texttt{Swept} implementation studied here offers double the speedup
of the GPU-only case studied previously~\cite{OurJCP} for the same grid size.
The improvement in relative speedup of the heterogeneous swept rule is expected since
latency is much worse for internode communication than within the GPU memory hierarchy
or between the CPU and GPU.

Figure~\ref{fig:BestRuntimeEuler} compares the computational cost per time step of the
\texttt{Classic} and \texttt{Swept} schemes applied to the Euler equations; the performance trends match those of the heat equation. 
In this case, the communication costs that the program avoids are significant enough that \texttt{Swept} decomposition provides a tangible benefit of 1.1--\tx{2.0}
despite the extra complexity, management, and memory resources that it requires.
This shows that swept time-space domain decomposition is a viable method for solving complex equations on systems with substantial communication costs and various architectures.
This contrasts our prior GPU-only results~\cite{OurJCP} where the swept decomposition scheme
always performed slower than classic decomposition.

Furthermore, the \texttt{Swept} scheme is less sensitive to launch conditions, and offers a regular performance profile with increasing problem size, compared with the \texttt{Classic} scheme.
The regularity of the \texttt{Swept} program performance
allows us to present fitted results in Table~\ref{tab:tablefit}, corresponding to the data points presented in Figures~\ref{fig:BestRuntimeHeat} and ~\ref{fig:BestRuntimeEuler}, that may illuminate and
guide future work on this and similar topics.

\section{Conclusions}
\label{sec:hConc}

In this study, we examined the performance characteristics of design choices that must be made when applying the swept time-space decomposition rule to partial differential equations on heterogeneous computational architectures. 
These design choices we considered are: how to efficiently and generally store the working array throughout the simulation, how many threads per block---i.e., points per domain---to assign, and what proportion of the total domain to assign to a GPU. 

We aimed to answer the primary questions concerning these design choices laid out in Section~\ref{sec:obj1}. 
First, we found that the best number of grid points to assign to each domain varies with the algorithm, governing equation(s), and grid size. 
To achieve the best performance on repeated similar runs, any program should be tested over a limited number of time steps and tuned to the best result; however, we recommend choosing a larger block size (e.g., \numrange{768}{1024}) that is a multiple of \num{32}. 
Next, we concluded the swept scheme appears relatively insensitive to the share of work given to the GPU, though maximum performance is achieved by tuning based on the problem size and number of threads per block.
For more complex problems (e.g., the Euler equations), we found better performance for GPU work factors of \numrange{50}{75}.

While tuning for optimal performance of the swept rule is highly problem dependent, we can suggest a few qualitative heuristics based on the optimal performance in Figures~\ref{fig:HeatContoursFull} and~\ref{fig:EulerContoursFull}. 
In terms of threads per blocks, a marginal majority suggests that using
larger numbers of threads leads to better performance, i.e., when choosing a range for your tuning experiment start with more and reduce as necessary. 
In terms of GPU work factor, the results also marginally suggest that larger values perform better, though the swept rule seems insensitive to this parameter overall when assigning most work to the GPU.

Furthermore, there is a significant tradeoff between extensibility and performance associated with the primary data structure and compression scheme applied to the working quantity in the simulation. 
The \texttt{lengthening} approach offers both performance benefits to the swept decomposition scheme as well as simplified development. 
Finally, although any conclusions drawn from an experiment on only two compute nodes are limited, we showed an improvement over previous results for solving the Euler equations using a fine-tuned GPU-only program~\cite{OurJCP}.

Future work should focus on adapting the swept rule to higher dimensions, architecture types, and grid formations. 
For example, while Alhubail and Wang demonstrated the two-dimensional swept rule for CPU-based clusters~\cite{Alhubail:2016arxiv}, we have not yet extended this to heterogeneous systems. 
In addition, we recognize the need to develop new experiments that examine the scaling characteristics of the program as additional computational resources are added. 
Additional experiments should be performed on cloud systems like Amazon Web Services, Microsoft Azure, or Nvidia GPU Cloud. 
These will provide greater insight into the factors affecting performance and help develop a more robust
performance model for swept time-space decomposition.

\section*{Acknowledgements}

This material is based upon work supported by NASA under award No.~NNX15AU66A
under the technical monitoring of Drs.~Eric Nielsen and Mujeeb Malik.
We also gratefully acknowledge the support of NVIDIA Corporation, who donated a
Tesla K40c GPU used for this research.

\appendix

\renewcommand*{\thesection}{\appendixname~\Alph{section}}

\section{Availability of material}
\label{app:material}

The software package \texttt{hSweep v2.0} used to perform this study is available
openly~\cite{hSweepz}; the most recent version can be found at its GitHub repository
shared under an MIT License: \url{https://github.com/Niemeyer-Research-Group/hSweep}.
All figures, and the data and plotting scripts necessary to reproduce them, for
this article are available openly under the CC-BY license~\cite{repropack}.

\section*{References}

\bibliographystyle{elsarticle-num}
\bibliography{paper-heterogeneous-swept}

\end{document}